\documentclass[conference]{IEEEtran}
\IEEEoverridecommandlockouts

\usepackage{cite}
\usepackage{amsmath,amssymb,amsfonts}
\usepackage{siunitx}
\usepackage{graphicx}
\usepackage{textcomp}
\usepackage{xcolor}
\usepackage{booktabs}
\usepackage{orcidlink}
\usepackage{soul}
\usepackage{tikz}
\usepackage{eso-pic}
\usetikzlibrary{arrows.meta,calc,decorations.pathmorphing}

\def\BibTeX{{\rm B\kern-.05em{\sc i\kern-.025em b}\kern-.08em
    T\kern-.1667em\lower.7ex\hbox{E}\kern-.125emX}}

\newcommand{\ampA}{GA-HBT-1}   
\newcommand{\ampB}{GA-HBT-2}  
\newcommand{\ampC}{GA-MMIC-1}  
\newcommand{\ampD}{GA-MMIC-2}  
\newcommand{\ampE}{GA-MMIC-3}  
\newcommand{\ampF}{IGP-HBT-1}  
\newcommand{\ampG}{IGP-HBT-2}  
\newcommand{\ampH}{GA-HBT-3}   
\newcommand{\ampI}{GA-HBT-4}   
\newcommand{\ampJ}{GA-pHEMT}   
\newcommand{\ampK}{GA-HBT-5}   
\newcommand{\ampL}{IGP-HBT-3}  

\begin{document}

\title{Residual AM and PM Noise of Selected\\ X-Band Amplifiers}
\author{%
  \IEEEauthorblockN{Archita Hati\orcidlink{0000-0001-8363-8648}\textsuperscript{1},
    Craig Nelson\orcidlink{0000-0001-5293-5113}\textsuperscript{1,2},
    Scott Littrell\textsuperscript{3}}
  \IEEEauthorblockA{%
    \textsuperscript{1}Time and Frequency Division, National Institute of Standards and Technology, Boulder, CO 80305 USA\\
    \textsuperscript{2}Frequency Electronics Inc., Boulder, CO 80301 USA\\
    \textsuperscript{3}KBM Enterprises Inc., Huntsville, AL 35805 USA\\
    Corresponding author: Archita Hati (email: archita.hati@nist.gov).}
}
\maketitle

\begin{abstract}

In microwave systems design, amplitude modulation (AM) noise is often ignored relative to phase modulation (PM) noise. However, in high-performance applications requiring ultra-low PM noise, AM noise becomes a critical limiting factor due to AM-to-PM conversion mechanisms. Because AM noise specifications are rarely, if ever, provided for commercial microwave components, this study characterizes the residual AM and PM noise of several 10~GHz amplifiers. We present performance data for various Heterojunction Bipolar Transistor (HBT) amplifier designs and demonstrate the noise reduction achieved through power-summed configurations. For example, a discrete-die GaAs HBT amplifier (labeled \ampA) achieves residual PM noise below $-160$~dBrad$^2$/Hz and AM noise of approximately $-155$~dB/Hz at a 1~kHz offset in a quad power-summed configuration. A notable finding is that the majority of amplifier designs surveyed in this study exhibit higher AM noise relative to their PM noise, with only a few exceptions. The impact of bias supply quality on residual noise is also quantified, demonstrating that the choice of voltage regulator is a first-order design parameter below 1~kHz offset.
\end{abstract}


\section{Introduction}

Residual amplitude modulation (AM) and phase modulation (PM) noise are fundamental metrics that set the noise floor of radar waveform generators, high-speed communication transmitters, and precision frequency synthesis systems~\cite{richards_principles_2010, melvin_principles_2012, charvat_small_2014, lathi_modern_1983,tomasi_advanced_2003, harris_multirate_2004, khanzadi_calculation_2014, piemontese_phase_2021, hati_state---art_2013, hati_radio_2025}. In the design community, residual PM noise has been extensively studied and characterized over several decades~\cite{aramburo_comparison_1997, driscoll_phase_1990, rubiola_phase_1992, mossammaparast_phase_2000, breitbarth_additive_2008, barnes_residual_2011,  boudot_phase_2012, hati_ultra-low-noise_2012, hati_pm_2014, rubiola_companion_2023}. Residual AM noise, however, has received substantially less attention in the primary literature~\cite{ascarrunz_investigations_1993, nelson_relationship_1994, walls_origin_1997, hati_comparison_2006, rubiola_effect_2007, calosso_phase-noise_2020, badillo_experimental_2022} and is almost universally absent from commercial component datasheets.

In any component exhibiting even modest amplitude-to-phase nonlinearity, AM noise ($S_\alpha(f)$) at the input is converted to PM noise ($S_\varphi(f)$) at the output through the AM-to-PM conversion.  This mechanism is particularly consequential in frequency multiplication chains, where AM noise from an amplifier stage enters a subsequent multiplier or mixer and reappears as PM noise at the multiplied output.

The practical consequences for phase noise metrology were documented by Cibiel et al.~\cite{cibiel_am_2002}, who showed that AM noise from amplifiers in a cross-spectrum measurement bridge can directly corrupt PM noise floors below $-167$~dBrad$^2$/Hz. Rubiola and Boudot~\cite{rubiola_effect_2007} extended this analysis to quantify the cross-talk floor introduced by finite AM rejection in homodyne detector bridges. The present work builds on the prior study by Hati, Howe, and Nelson~\cite{hati_comparison_2006}, which characterized both AM and PM noise in X-band oscillators and amplifiers circa 2006. The intervening years have seen the introduction of a new generation of Gallium Arsenide
Heterojunction Bipolar Transistor (GaAs HBT)  and Monolithic Microwave Integrated Circuit (MMIC) devices; whether they represent a substantive improvement in AM noise performance over legacy devices is an open and practically important question. With the wider availability of commercial phase noise analyzers, more manufacturers now provide PM noise information in their datasheets, but to the best of our knowledge AM noise specifications remain universally absent.

The present study was motivated directly by the stringent PM and AM noise requirements of a frequency synthesis development program at National Institute of Standards and Technology (NIST). Identifying X-band amplifier designs that meet or exceed the noise budget requires measured residual AM and PM data that does not exist in any manufacturer documentation. The scope is: (a) 10~GHz center frequency; (b) mostly GaAs HBT technology (discrete die, MMIC, and module-level packages); (c) single, dual, and quad power-summed array configurations; and (d) characterization across a range of input drive levels up to and including mild compression. The central questions are which amplifier designs and noise mechanisms yield the lowest combined residual AM and PM noise, and what improvements in AM and PM noise can be achieved with array-summing.

\section{Measurement System and Methodology}
\label{sec:measurement}

\subsection{System Architecture}

\begin{figure}[htbp]
  \centering
  \includegraphics[width=\columnwidth]{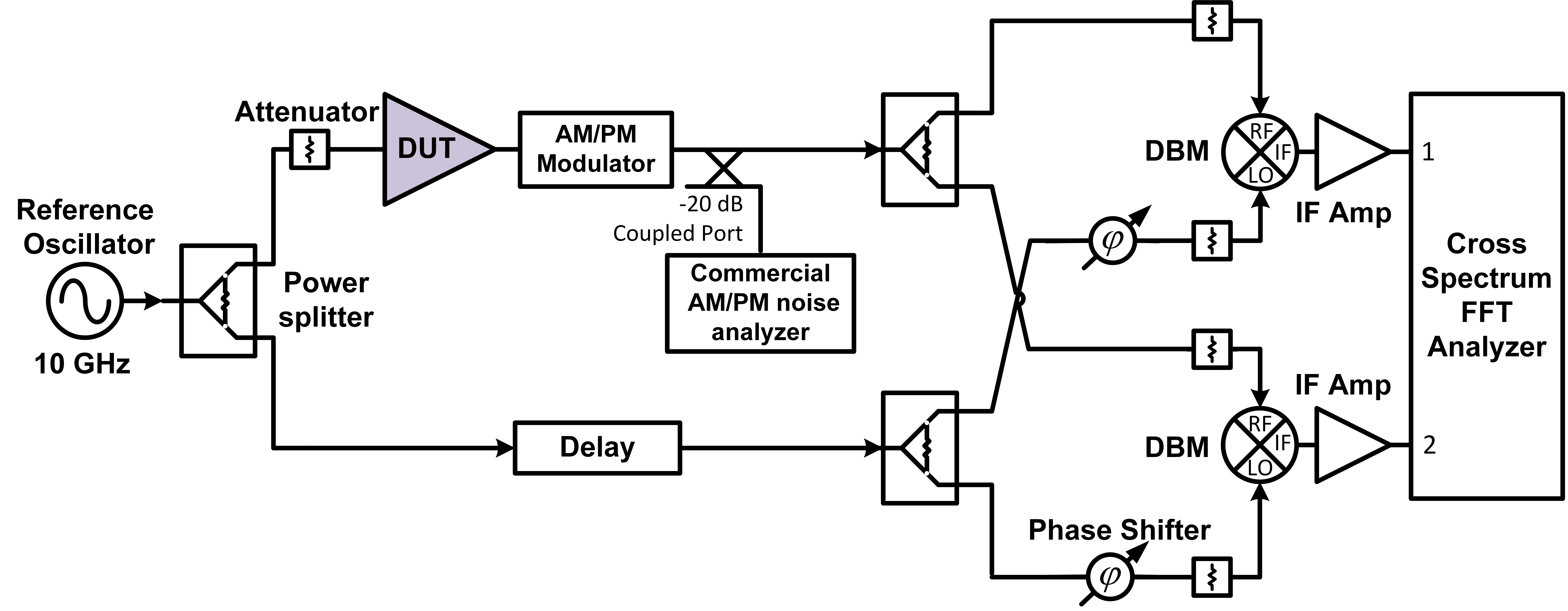}
  \caption{Block diagram of the custom-built cross-spectrum measurement system for residual AM and PM noise.
    DUT: Device Under Test, FFT: Fast Fourier Transform, DBM: Double Balanced Mixer. The DBM acts as either a phase or amplitude detector. }
  \label{fig:measurement_setup}
\end{figure}

Figure~\ref{fig:measurement_setup} shows the cross-spectrum homodyne measurement system, referred to hereafter as the NIST phase noise analyzer (NIST PNA), used throughout this study~\cite{walls_cross-correlation_1992, rubiola_cross-spectrum_2010, nelson_phase_2013}. The cross-spectrum technique is employed because it allows the system noise floor to be suppressed below the residual noise of the device under test (DUT) by cross-correlating the outputs of two independent measurement channels.

Let $x(t)$ and $y(t)$ denote the outputs of channels~1 and~2,
respectively, and let $X(f)$ and $Y(f)$ denote their Fourier transforms. Each channel output can be decomposed as
\begin{equation}
  x(t) = d(t) + n_x(t), 
  \qquad
  y(t) = d(t) + n_y(t)
  \label{eq:channel_model}
\end{equation}

\noindent where $d(t)$ represents the common DUT noise contribution and
$n_x(t)$ and $n_y(t)$ are the uncorrelated noise contributions of the individual channels (mixer, IF amp, etc.). Capitol letters represent the Fourier transform of the respective lowercase variables. Under the assumption that $n_x$, $n_y$, and $d$ are mutually uncorrelated, the cross-power spectral density $S_{xy}$ is estimated as the ensemble average, over $m$ acquisitions, of the product $X(f)Y^*(f)$, where $^*$ denotes complex conjugation \cite{rubiola_cross-spectrum_2010}:

\begin{equation}
\begin{split}
 \left\langle S_{xy} \right\rangle_m 
 &= {\frac{2}{T}} \left\langle X(f)Y^*(f) \right\rangle_m
 \\&={\frac{2}{T}}\left\langle D(f)D^*(f)+D(f)N^*_y(f)\right. \\
&\quad \left. +N_x(f)D^*(f) +N_x(f)N^*_y(f)\right\rangle_m
 \\&=\ S_d(f)_{m\rightarrow\infty}.
 \end{split}
\end{equation}
\noindent Here, $T$ is the measurement or truncation time normalizing the power spectral density (PSD) to a 1~Hz bandwidth, and $S_d(f)$ is the one sided PSD of the common DUT noise $d(t)$. This means that the cross-spectrum estimator converges to the true DUT noise PSD, as the cross-terms average to zero as the number of averages $m \to \infty$. In practice the residual channel noise scales as $1/\sqrt{m}$, enabling characterization of devices whose residual noise would otherwise be masked by the measurement noise floor. 
Unlike residual PM noise measurements, where reference oscillator PM noise cancels to a high degree, AM noise does not. Therefore, valid residual AM noise measurements require the source AM noise to be substantially below that of the DUT (see Appendix~\ref{app:residual_noise}). In the present configuration (Fig.~\ref{fig:measurement_setup}), the cross-spectrum floor is confirmed to be 5~dB to 15~dB below the residual noise of the lowest-noise device under test. Furthermore, to minimize the anti-correlated thermal noise contribution of the power splitters \cite{hati_cross-spectrum_2016}, it is important not to attenuate the signal at the splitter input, and rather set the detector power level directly prior to the AM/PM detector as shown in Fig.~\ref{fig:measurement_setup}. 

\subsection{Reference Oscillator}
\begin{figure}[htbp]
  \centering
  \includegraphics[width=.6\columnwidth]{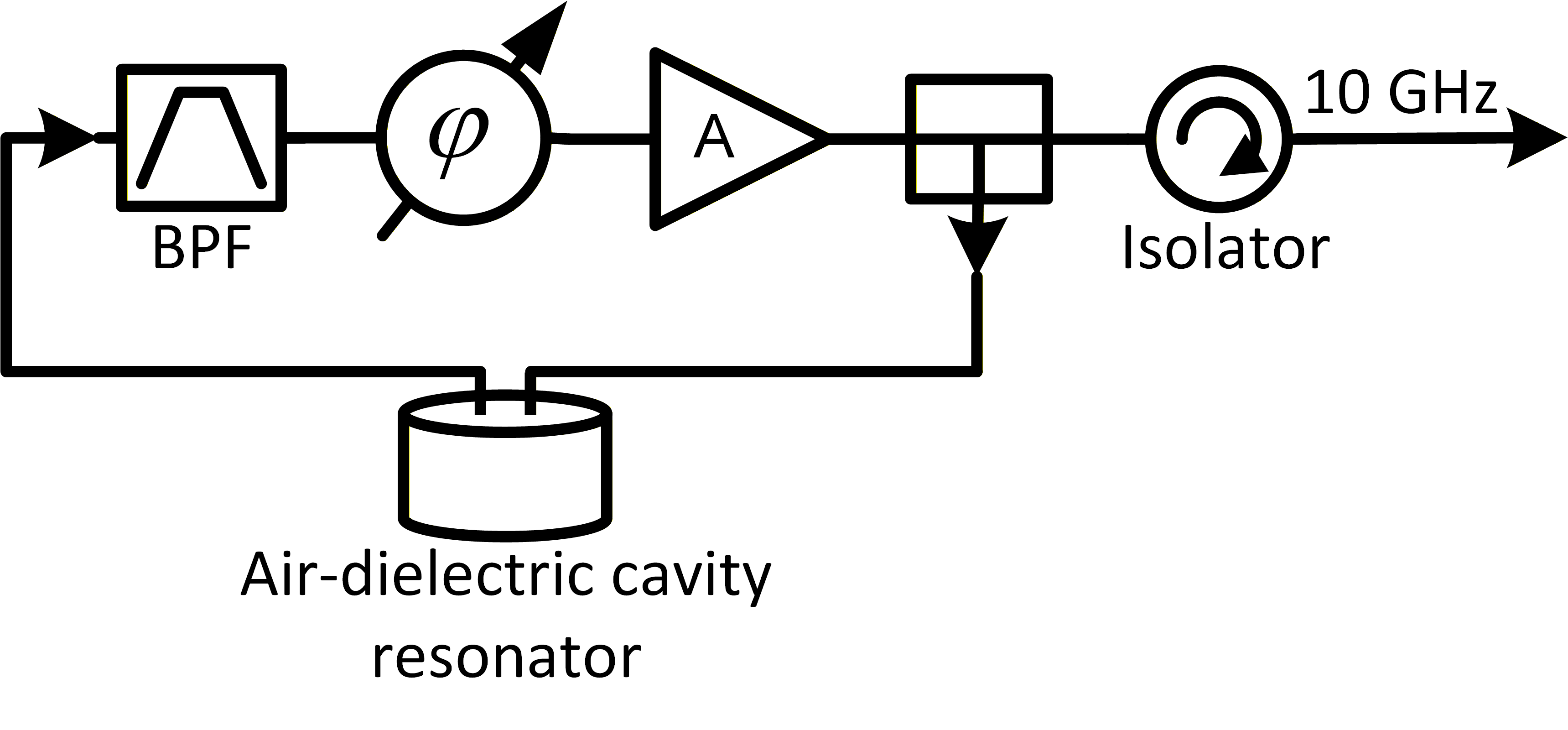}
  \caption{Block diagram of the simple loop oscillator used as a reference frequency source. BPF - Bandpass filter, A - Low AM noise loop amplifier.}
  \label{fig:Sloop}
\end{figure}

\begin{figure}[htbp]
  \centering
  \begin{tikzpicture}
    \node[anchor=south west,inner sep=0] (img) at (0,0)
      {\includegraphics[width=\columnwidth]{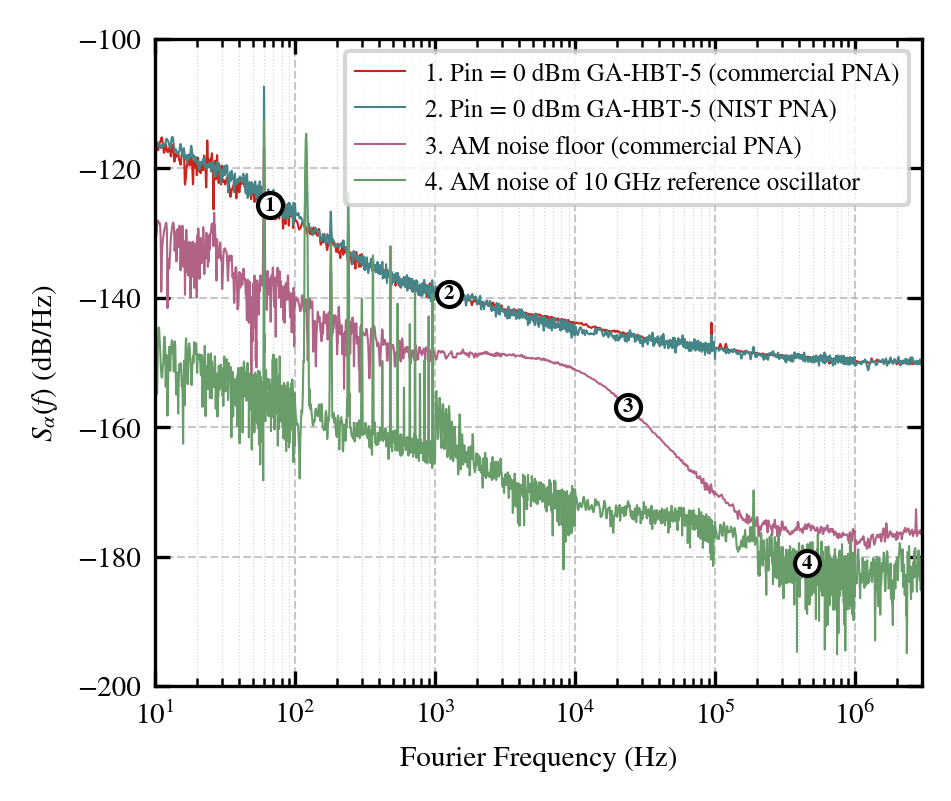}};
    \begin{scope}[x={(img.south east)},y={(img.north west)}]
      \coordinate (o) at (0.185,0.255);
    \end{scope}
    \begin{scope}[shift={(o)},x=0.85mm,y=0.85mm,line width=0.4pt,
                  every node/.style={font=\tiny\bfseries,inner sep=0.25mm},
                  >={Latex[length=0.9mm,width=0.8mm]}]
      \draw[fill=white] (3,0) circle[radius=2.7];
      \draw[line width=0.35pt] plot[domain=-1.9:1.9,samples=40,smooth] (\x+3,{1.4*sin(\x*95)});
      \node[anchor=north] at (3,-3.1) {10 GHz};
      \draw[->] (5.7,0) -- (8,0);
      \draw[fill=white] (8,-2) rectangle (12,2);
      \draw[line width=0.35pt,decorate,decoration={zigzag,amplitude=0.35mm,segment length=0.8mm}] (8.8,0) -- (11.2,0);
      \node[anchor=south] at (10,2.2) {A1};
      \draw[->] (12,0) -- (14.3,0);
      \draw[fill=blue!14] (14.3,-3.5) -- (14.3,3.5) -- (21,0) -- cycle;
      \node at (16.5,0) {DUT};
      \draw[->] (21,0) -- (23.3,0);
      \draw[fill=white] (23.3,-2) rectangle (27.3,2);
      \draw[line width=0.35pt,decorate,decoration={zigzag,amplitude=0.35mm,segment length=0.8mm}] (24.1,0) -- (26.5,0);
      \node[anchor=south] at (25.3,2.2) {A2};
      \draw[->] (27.3,0) -- (29.6,0);
      \draw[fill=white] (29.6,-3.5) rectangle (46.5,3.5);
      \node[align=center] at (38.05,0) {AM/PM noise\\analyzer};
    \end{scope}
  \end{tikzpicture}
  \caption{Comparison of AM noise measurements for amplifier \ampK\ using a commercial PNA and the NIST PNA. (i) Traces 1 and 2: Excellent agreement between the two systems for the DUT. (ii) Trace 3: AM noise of the 10 GHz reference measured directly by the commercial PNA (DUT and attenuator A2 bypassed), illustrating the instrument's internal AM noise floor at 10 GHz. (iii) Trace 4: Residual AM noise of the 10~GHz reference oscillator (DUT bypassed), measured via the Fig.~\ref{fig:measurement_setup} configuration.}
  \label{fig:AM_verification}
\end{figure}

The system reference is an air-dielectric cavity resonator oscillator in a simple loop configuration operating at 10~GHz (Fig.~\ref{fig:Sloop}).  This oscillator exhibits an exceptionally low AM noise level, as illustrated in Fig.~\ref{fig:AM_verification}, $S_\alpha(\SI{1}{\kilo\hertz}) \approx -162$~dB/Hz, which was set by an obsolete commercial eight-element HBT array amplifier (\ampL, Table~\ref{tab:dut}) operating in deep compression of about 7~dB. This performance ensures that the reference source AM noise in the flicker region is well below the residual AM noise of the amplifiers selected for this study.

\subsection{AM/PM Detector Configuration}

Rather than employing conventional diode-based AM detectors \cite{walls_measurement_1999, rubiola_measurement_2005, rubiola_effect_2007}, this system uses double-balanced mixers (DBMs) as homodyne detectors\cite{rubiola_tutorial_2006}. The primary reason for this configuration was the ease of switching between AM and PM measurements with the same set-up. The mixer sensitivity to AM or PM is selected by setting the relative phase between the local oscillator (LO) and radio frequency (RF) ports of each mixer: $0^\circ$ phase yields AM detection, $90^\circ$ phase yields PM detection. For PM detection both mixer ports are driven at or above saturation, whereas for AM detection the RF port power is backed off into its linear regime. The AM-to-PM and PM-to-AM leakage and its dependence on saturation are analyzed in Appendix~\ref{app:residual_noise}. An integrated AM/PM modulator provides the reference modulation tone required for absolute calibration and alignment of both channels. Cross-talk between the AM and PM sensitivity, arising from imperfect phase alignment, is suppressed by adjusting the phase shifter until the orthogonal alignment tone nulls in the detector channel.

\subsection{Verification Instrument}

A commercial cross-correlation phase noise analyzer (PNA) with AM noise measurement capability was connected as a monitor channel for all measurement setups. It enables quick and convenient alignment of the modulator for pure AM and PM tones. The commercial PNA is not used as the primary measurement instrument; the NIST PNA (Fig.~\ref{fig:measurement_setup}) provides the reported AM and PM noise spectral densities. However, the commercial PNA was occasionally used to verify the accuracy of the AM and PM measurement. Although the AM noise floor of the commercial PNA was insufficient for characterizing some of the low-noise amplifiers, it allowed for the validation of analog measurements on a higher-noise device as shown in Fig.~\ref{fig:AM_verification}.

\subsection{Calibration Procedure}

Calibration of residual AM and PM noise spectral density is performed by injecting a known modulation from the integrated AM/PM modulator~\cite{armstrong_method_1936, sann_measurement_1968, nelson_extending_1996, ivanov_generation_2012, hati_simple_2022,hati_ultralow_2024} at multiple offset frequencies and recording the detector output level. The modulation index is set to a known value, establishing the detector transfer coefficients $k_\varphi$ in V/rad for PM and $k_\alpha$ in V for AM. This calibration is performed at each input power setting for all amplifiers to account for any operating-point dependence in the mixer conversion gain. Attenuators at the RF input of the mixers were adjusted for each amplifier to achieve roughly constant conversion factors across all PM measurements; this same procedure was repeated for AM measurements, albeit at a different mixer operating point.

Accurate $k_\alpha$ calibration necessitates placing the AM/PM modulator after the DUT in Fig.~\ref{fig:measurement_setup}. Because AM noise measurements require the mixer’s RF port to operate strictly within its linear regime, this specific placement prevents compression of the AM calibration tone which would otherwise occur if the modulator is placed in the saturated reference arm, yielding erroneous $k_\alpha$ values. Conversely, for PM noise and $k_\varphi$ calibration, the modulator can be positioned either on the reference arm or post-DUT.

\subsection{Array Configuration}

This study evaluates the noise performance of both commercial and custom-built amplifier arrays. All custom  power-summed architectures utilize Wilkinson power dividers and combiners to distribute and recombine signals across $n$ parallel paths. For an array of $n$ amplifiers driven by split copies of a common 
input signal, the summed output amplitude and phase noise spectra 
are given by \cite{boudot_phase_2012, pomponio_ultra-low_2024}
\begin{equation}
    S_{\alpha}(f) = \frac{S_{\alpha-f}(f)}{n} + 
    \frac{N_F k_B T}{P_{in}}
\end{equation}
\begin{equation}
    S_{\varphi}(f) = \frac{S_{\varphi-f}(f)}{n} + 
    \frac{N_F k_B T}{P_{in}}
\end{equation}
where $S_{\alpha-f}$ and $S_{\varphi-f}$ are the flicker amplitude  and phase noise of a single amplifier, $N_F$ is the noise figure,  $k_B T$ is the thermal noise power in 1~Hz bandwidth ($\approx -174$~dBm/Hz at room temperature), and $P_\mathrm{in}$ is the input power to the array amplifier. The multiplicative flicker noise is reduced by  the array factor $n$ because the correlated carrier amplitudes add  linearly while the uncorrelated flicker noise variables add in  power. The additive thermal noise, though also uncorrelated  across amplifiers and thus subject to the same $1/n$ power  averaging, receives no net benefit because the input splitter  reduces the signal power delivered to each amplifier by the same  factor $n$, leaving the thermal noise unchanged relative to  a single amplifier.

\begin{figure}[htbp]
  \centering
  \includegraphics[width=0.75\columnwidth]{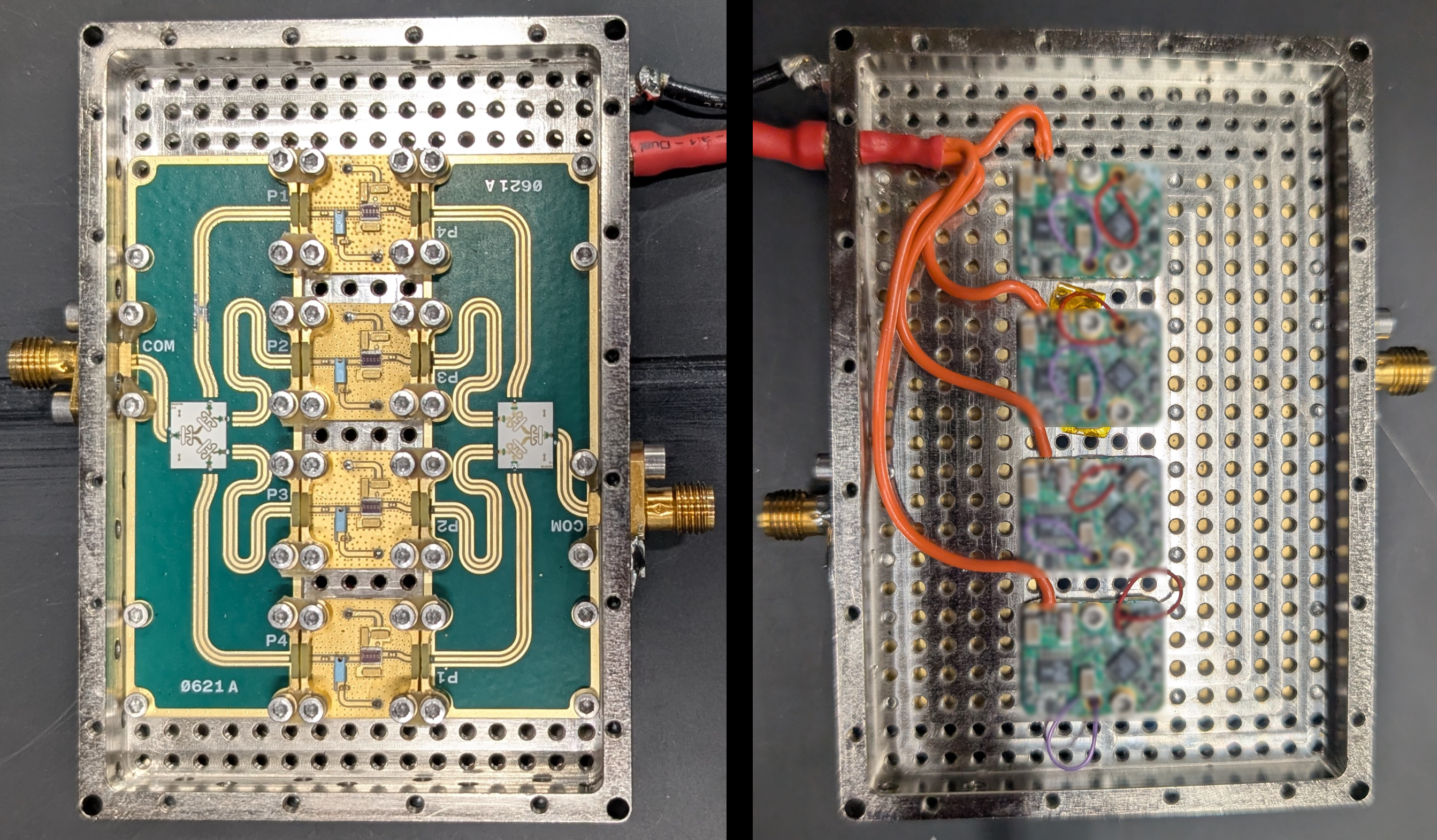}
  \caption{Picture of the custom-built quad array of amplifier \ampD. Four voltage regulators (right), directly mounted underneath the amplifiers, provide independent bias voltage to each element of the array (left).}
  \label{fig:quad_array_photo}
\end{figure}

In an ideal case of an amplifier array, the gain and noise figure remain the same while the 1~dB compression point increases. In practice, several factors cause performance to deviate from this theoretical limit. First, the excess insertion loss of the Wilkinson power splitters attenuates the signal, thereby reducing gain, the thermal signal-to-noise ratio (SNR), and total power output. Second, imbalances in amplitude and phase between channels prevent perfect coherent recombination, further degrading gain and inducing AM$\leftrightarrow$PM conversion. Additionally, unit-to-unit variations in amplifier parameters and noise correlation, often arising from shared power supplies or environmental coupling, preclude the ideal $10\log_{10}(n)$ improvement. To mitigate correlated power supply noise, which can limit the statistical advantage of uncorrelated noise addition of the array, individual voltage regulators were used for each amplifier. As shown in Fig.~\ref{fig:quad_array_photo}, the custom-built array consists of four amplifiers, with each element utilizing an independent regulation stage to ensure maximum isolation and noise performance.

For all array amplifiers discussed, the input power ($P_\mathrm{in}$) refers to the total power at the splitter input, rather than the power entering each individual element of the array. 

\begin{table*}[!t]
  \caption{Amplifiers Characterized in This study}
   \label{tab:dut}
  \centering
  \footnotesize
  \renewcommand{\arraystretch}{1.25}
  \begin{tabular}{llllcccccc}
    \toprule
        \textbf{Device} & \textbf{Package /} & \textbf{Technology}
      & \textbf{BW} & \textbf{Gain} & \textbf{P1dB}
      & \textbf{OIP3} & \boldmath{$N_F$} & {\boldmath $S_\varphi(\textbf{1~kHz})$}
      & {\boldmath $S_\alpha(\textbf{1~kHz})$} \\
        \textbf{label}       & \textbf{topology} & \textbf{(from datasheet)}
      & \textbf{(GHz)} & \textbf{(dB)} & \textbf{(dBm)}
      & \textbf{(dBm)} & \textbf{(dB)} & \textbf{(dBrad$^2$/Hz)} & \textbf{(dB/Hz)}\\
    \midrule
    \ampA\ (S,Q)  & Discrete die        & GaAs HBT        & 2--30  & 12   & 21 & 33 & 5.0 & $-157$ & $-149$ \\
    \ampB\ (S,Q)  & Discrete die        & GaAs HBT        & 6--12  & 15.6 & 20 & 32 & 5   & $-153$ & $-147$ \\
    \ampC\ (S,Q)  & Packaged MMIC       & GaAs MMIC       & 2--20  & 15   & 20 & 31 & 3.2 & $-150$ & $-142$ \\
    \ampD\ (S,Q)  & Packaged MMIC       & GaAs MMIC       & DC--18 & 14.4 & 24 & 34 & 3.5 & $-150$ & $-143$ \\
    \ampE\ (S)    & Packaged MMIC       & GaAs MMIC       & 6--18  & 18   & 20 & 29 & 4.5 & $-152$ & $-143$ \\
    \ampF\ (D)    & Array (2)           & InGaP/GaAs HBT  & DC--12 & 9.7  & 13 & 23 & 4.9 & $-150$ & $-149$ \\
    \ampG\ (S,Q)  & Distributed die     & InGaP/GaAs HBT  & 2--18  & 14   & 15 & 27 & 5   & $-152$ & $-145$ \\
    \ampH\ (S)    & Hermetic module     & GaAs HBT        & 6--12  & 11   & 23 & 34 & 4.5 & $-156$ & $-152$ \\
    \ampI\ (S)    & Hermetic module     & GaAs HBT        & 7--11  & 9    & 22 & 33 & 6   & $-158$ & $-155$ \\
    \ampJ\ (S)    & Distributed         & GaAs pHEMT      & DC--22 & 13.5 & 28 & 39 & 3.5 & $-152$ & $-140$ \\
    \ampK\ (S)    & Connectorized\ module       & GaAs HBT        & 2--30  & 11   & 20 & 29 & 4.5 & $-152$ & $-140$ \\
    \ampL\ (O)    & Array (8)           & InGaP HBT       & 8--12  & 18   & 21 & 29 & 8   & $-159$ & $-155$ \\
    \bottomrule
    \multicolumn{10}{l}{\scriptsize S = Single, D = Dual, Q = Quad, O = Octa. Devices are anonymized; labels denote package/technology class only.}\\ 
    \multicolumn{10}{l}{\scriptsize Bandwidth (BW), Gain, Noise Figure ($N_F$), 1-dB compression point (P1dB), Output Third-Order Intercept Point (OIP3) at 10~GHz.}  \\
    \multicolumn{10}{l}{\scriptsize AM and PM noise are for the single amplifier configurations for a $P_\mathrm{in}$ = 0~dBm.} \\
    \multicolumn{10}{l}{\scriptsize \ampF\ and \ampL\ are measured as array configurations for a $P_\mathrm{in}$ = 0~dBm. } \\
    \multicolumn{10}{l}{\scriptsize AM and PM noise are measured at NIST, while all other parameters are sourced directly from the manufacturer's datasheet.}\\    
       \end{tabular}
\end{table*}

\section{Devices Under Test}
\label{sec:dut}

Table~\ref{tab:dut} lists all amplifiers characterized in this study. In accordance with NIST policy on the evaluation of commercial products, each device is identified only by a label that encodes its semiconductor technology and package or topology class; manufacturers and part numbers are not given, and the order of presentation does not imply a ranking. The devices were sourced from five manufacturers and span several Gallium Arsenide Heterojunction Bipolar Transistor (GaAs HBT) subtypes. \ampA\ and \ampB\ are discrete GaAs HBT die in chip carrier packages. \ampC, \ampD, and \ampE\ are packaged devices from a common product family and are listed by the manufacturer as GaAs MMICs. \ampF\ is an Indium gallium phosphide (InGaP)/GaAs HBT, supplied and measured as a two-element array. \ampG\ is a GaAs InGaP HBT MMIC distributed amplifier die. In contrast, \ampH\ and \ampI\ are GaAs HBT ultra-low phase noise amplifiers housed in hermetically sealed modules; both are now obsolete. \ampJ\ is a GaAs pseudomorphic high-electron-mobility transistor (pHEMT) distributed amplifier and is included for comparison with the HBT devices despite its different technology; it represents a distinct noise mechanism baseline. \ampK\ is a broadband GaAs HBT amplifier in a connectorized package. \ampL, though now obsolete, consists of an array of eight parallel amplifiers utilizing InGaP HBT technology. The technical data in Table~\ref{tab:dut} are based on available manufacturer documentation. Experimental results indicate performance variations between units of the same model, particularly for \ampC.
The amplifiers integrated into the quad arrays were distinct units from those utilized in the single-amplifier characterization measurements.

\section{Results}
\label{sec:results}

\subsection{Single-Amplifier PM Noise Comparison}

\begin{figure}[htbp]
  \centering
  \includegraphics[width=\columnwidth]{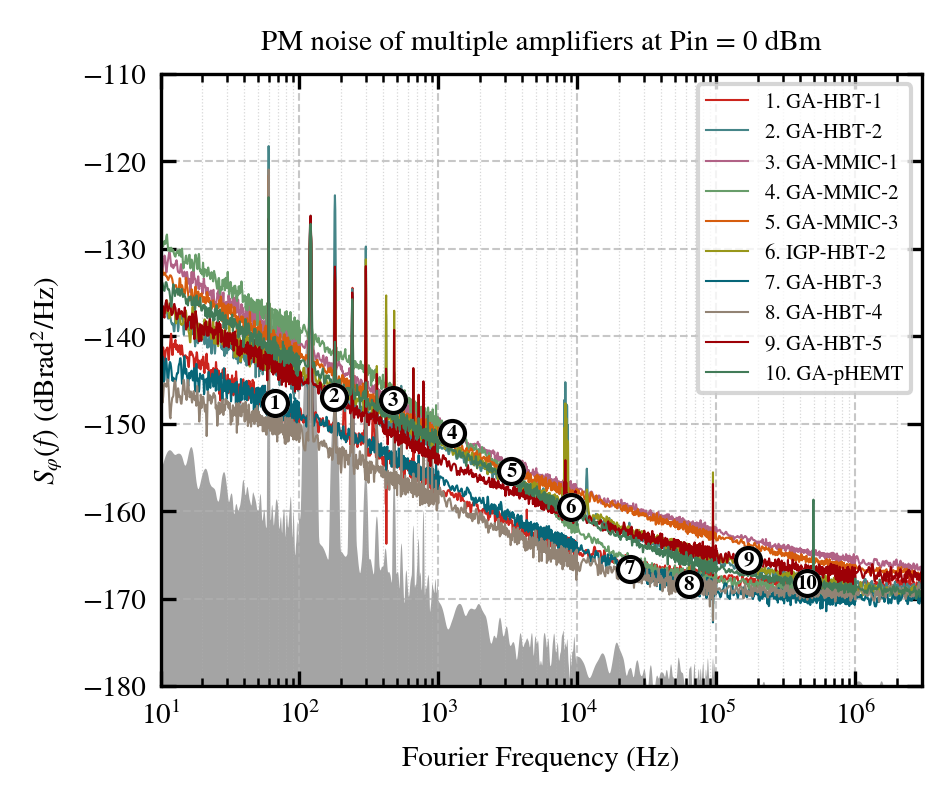}
  \caption{Measured PM noise of multiple amplifier designs (single configuration) at 10~GHz,
    $P_\mathrm{in} = 0$~dBm. Device details are in Table~\ref{tab:dut}; traces are numbered in table order, not by performance. Gray shaded area is the system PM noise floor for $P_\mathrm{in} = 0$~dBm. The number of FFT averages ($m$) for different frequency spans are as follows: 10~Hz to 100~Hz = 500, 100~Hz to 1~kHz = 1000, 1~kHz to 10~kHz = 3000, 10~kHz to 100~kHz = 5000, and 100~kHz to 3~MHz = 10,000.}
  \label{fig:Multi_Amp_PM}
\end{figure}

Figure~\ref{fig:Multi_Amp_PM} shows the residual PM noise of all characterized single amplifiers operating at input power, $P_\mathrm{in} = 0$~dBm. The spread in performance at 1~kHz offset is almost 10~dB across the full set and exceeds 15~dB at 10~Hz offset. Amplifier \ampA, together with the obsolete modules \ampH\ and \ampI, measured residual PM noise below approximately $-155$~dBrad$^2$/Hz at 1~kHz offset. 

\ampJ, which is a 1~W pHEMT rather than an HBT device, exhibits a residual PM noise floor that is comparable to the mid-power HBT devices despite its fundamentally different transport mechanism. The pHEMT flicker noise behavior is generally attributed to interface trap states at the channel heterointerface rather than to bulk recombination, and its measured PM noise is consistent with its high 1~dB compression point (P1dB) margin at the 0~dBm operating point.

\subsection{Single-Amplifier AM Noise Comparison}

\begin{figure}[htbp]
  \centering
  \includegraphics[width=\columnwidth]{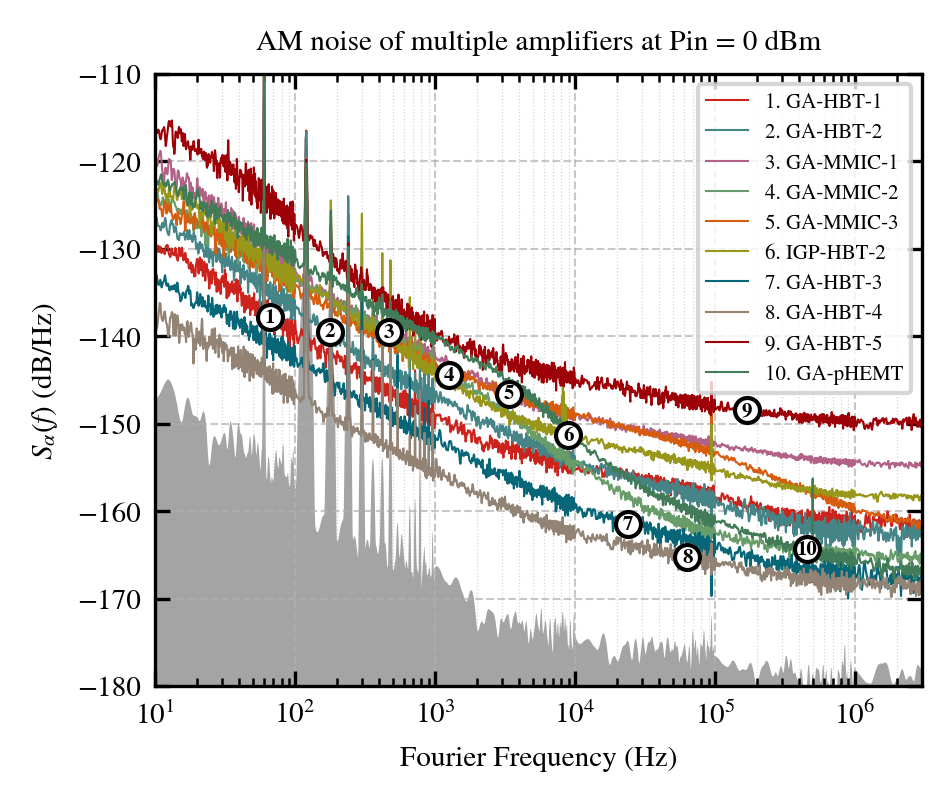}
  \caption{Measured AM noise of multiple amplifier designs (single configuration) at 10~GHz,
    $P_\mathrm{in} = 0$~dBm. Device details are in Table~\ref{tab:dut}; traces are numbered in table order, not by performance. Gray shaded area is the system AM noise floor for $P_\mathrm{in} = 0$~dBm. The number of FFT averages ($m$) for different frequency spans are as follows: 10~Hz to 100~Hz = 500, 100~Hz to 1~kHz = 1000, 1~kHz to 10~kHz = 3000, 10~kHz to 100~kHz = 5000, and 100~kHz to 3~MHz = 10,000.}
  \label{fig:Multi_Amp_AM}
\end{figure}

Figure~\ref{fig:Multi_Amp_AM} shows the corresponding residual AM noise. The performance
broadly follows the PM noise results, but two observations are
quantitatively significant.

First, the spread across devices is larger in AM than in PM.  Second, several devices show  AM noise 5~dB to 10~dB above their own PM noise in the flicker region below 1~kHz. Also, in the thermal region there is an assumption, widely held in system-level noise modeling, that AM noise is equal to the PM noise of the same device due to additive noise. However, as Fig.~\ref{fig:Multi_Amp_PM} and Fig.~\ref{fig:Multi_Amp_AM} show, the AM noise is higher than the corresponding PM noise by approximately 1~dB (\ampI) to 17.5~dB (\ampK) at 1~MHz offset, also indicating that the white AM noise mechanism is not additive.

\subsection{Measurement Uncertainty}
Rather than a full ``International Organization for Standardization - Guide to the Expression of Uncertainty in Measurement (ISO GUM)'' budget \cite{noauthor_evaluation_2008}, a streamlined approach is used to bound the dominant error terms of the AM and PM measurement system. The accuracy of the detector constants ($k_\varphi$ and $k_\alpha$) is verified using a NIST AM and PM white noise calibration standard \cite{hati_calibration_2012} with a 2-sigma uncertainty of $\pm$0.7~dB. Combining this uncertainty with the statistical variance from the $m = 1000$ cross-spectrum FFT averages yields a total 2-sigma uncertainty of $\pm$1.0~dB at a 1~kHz offset in Fig.~\ref{fig:Multi_Amp_PM} and Fig.~\ref{fig:Multi_Amp_AM}. 

AM-to-PM and PM-to-AM conversion terms are excluded from this budget because the measurement configuration suppresses each by more than 30~dB. For PM measurements, a pure AM tone is injected while mixer PM sensitivity is optimized, rejecting DUT AM noise at the RF port, and operating the mixer in saturation further reduces source AM noise. For AM measurements, the complementary configuration is used: a pure PM tone is injected while mixer AM sensitivity is optimized, rejecting DUT PM-to-AM conversion, and the source's common-mode PM noise cancels between the two ports. 

Finally, a worst-case noise floor bias of approximately +0.4~dB occurs at a 1~kHz offset on the lowest noise traces in Fig.~\ref{fig:Multi_Amp_PM} and Fig.~\ref{fig:Multi_Amp_AM}; this bias has been left uncorrected to maintain a conservative presentation of the raw data.

\subsection{\ampA: Array PM Noise Performance}

\begin{figure}[htbp]
  \centering
  \includegraphics[width=\columnwidth]{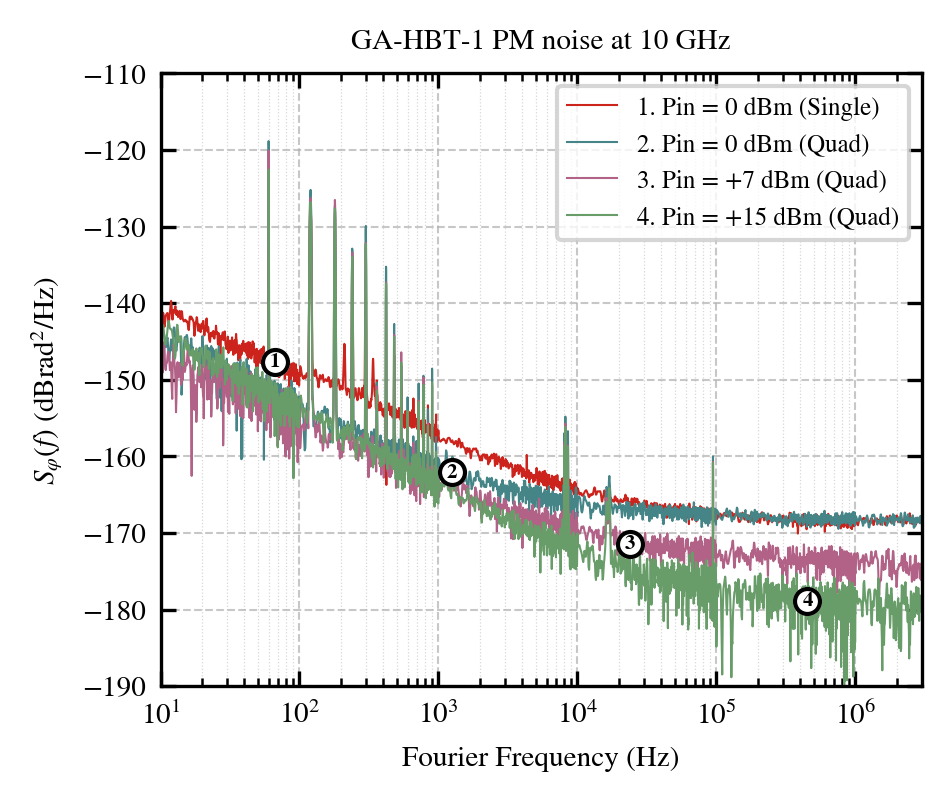}
  \caption{Residual PM noise of amplifier \ampA\ in single and quad
    configurations at 10~GHz.} 
  \label{fig:PM_ampA}
\end{figure}

Figure~\ref{fig:PM_ampA} shows the residual PM noise of \ampA\ (Discrete-Die HBT) in single and quad configurations at several input power levels. In the single configuration at $P_\mathrm{in} = 0$~dBm, \ampA\ achieves approximately $-157$~dBrad$^2$/Hz at 1~kHz offset. In the quad configuration, the flicker region improves by approximately 3~dB to 5~dB, reaching below $-161$~dBrad$^2$/Hz at 1~kHz offset. The varying improvement with offset frequency may be attributed to unit-to-unit
performance variations, correlation from power supply, etc. As expected and discussed earlier, the thermal white phase noise, $S_\varphi(f) \approx N_F k_B T / P_\mathrm{in}$ \cite{hati_noise_2003}, does not improve with the quad configuration at $P_\mathrm{in} = 0$~dBm relative to the single amplifier. However, for $P_\mathrm{in} = +7$~dBm, where per-element input power is roughly 0~dBm, a 5~dB to 6~dB improvement is observed at all offset frequencies relative to single amplifier at 0~dBm.

\subsection{\ampA: Array AM Noise Performance}
\label{sec:amp1_am}

\begin{figure}[htbp]
  \centering
  \includegraphics[width=\columnwidth]{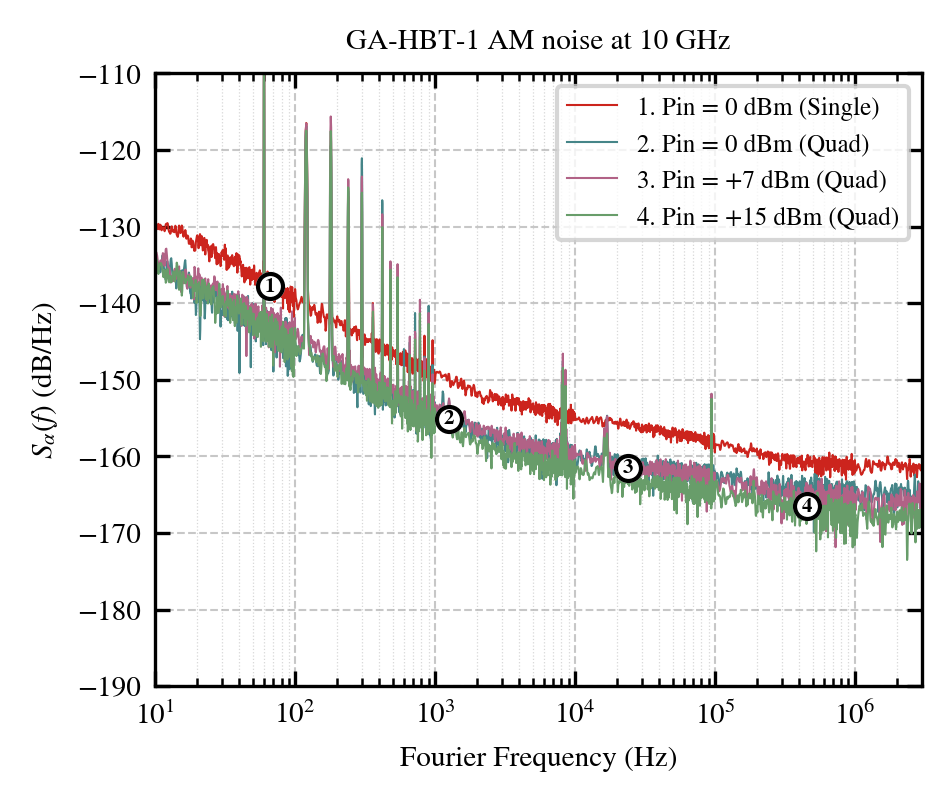}
  \caption{Residual AM noise of amplifier \ampA\ in single and quad
    configurations at 10~GHz. Note improvement in both flicker and
    white noise regions for the quad configuration.}
  \label{fig:AM_ampA}
\end{figure}

Figure~\ref{fig:AM_ampA} reveals the most technically significant result in this study. Unlike the PM noise, which shows an improvement in flicker but no change in the white phase noise, the AM noise of \ampA\ improves in both the flicker and white noise regions when converting operation from single to quad configuration for a fixed input power level. The observed improvement in the AM white noise is approximately 3~dB to 5~dB while the PM white noise remains unchanged (see  Fig.~\ref{fig:PM_ampA} and Fig.~\ref{fig:AM_ampA}, $P_\mathrm{in}$ = 0~dBm). This suggests that the white AM noise in these devices is dominated by internal excess noise rather than being set by the $50~\Omega$ SNR thermal limit. Additionally, in Fig.~\ref{fig:AM_ampA}, the three quad AM traces nearly overlap in the white region. This power independence provides stronger proof that the AM floor is not additive-thermal.

The amplitude noise of the single and quad \ampA\ is significantly higher than its phase noise across both the flicker and thermal frequency regions. Notably, in the quad configuration at an input power of +15~dBm, the difference between phase noise and amplitude noise exceeds 10 dB in the white noise region.

\subsection{\ampC: Array AM and PM Noise Performance}
\begin{figure}[htbp]
  \centering
  \includegraphics[width=\columnwidth]{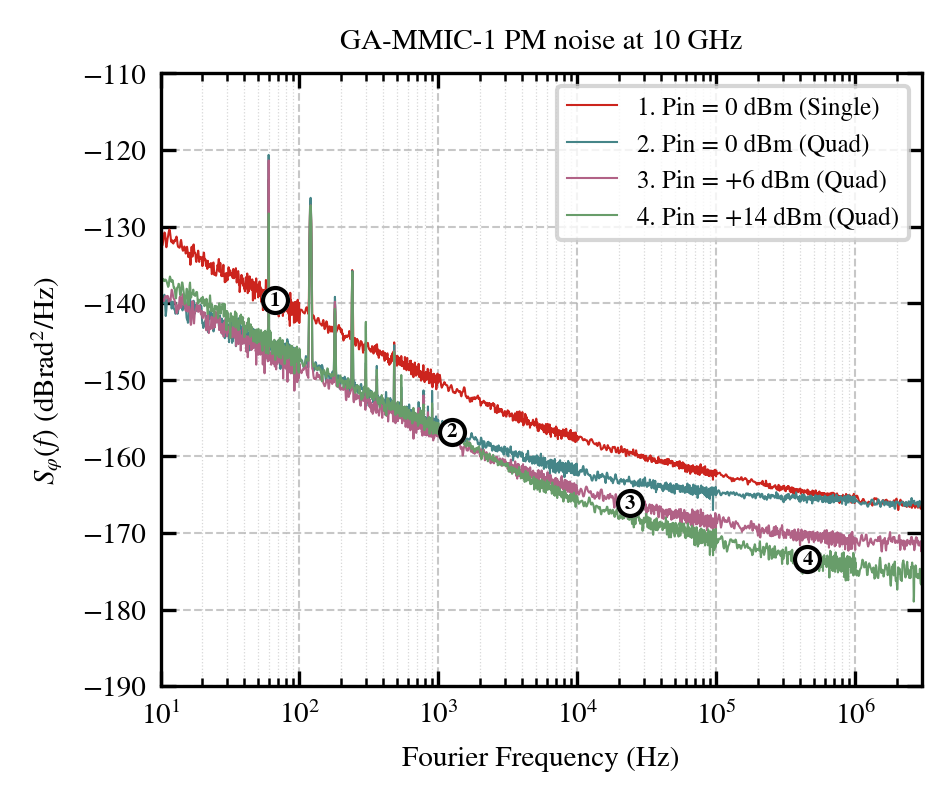}
  \caption{Residual PM noise of \ampC\ in single and quad
    configurations.}
  \label{fig:PM_ampC}
\end{figure}

\begin{figure}[htbp]
  \centering
  \includegraphics[width=\columnwidth]{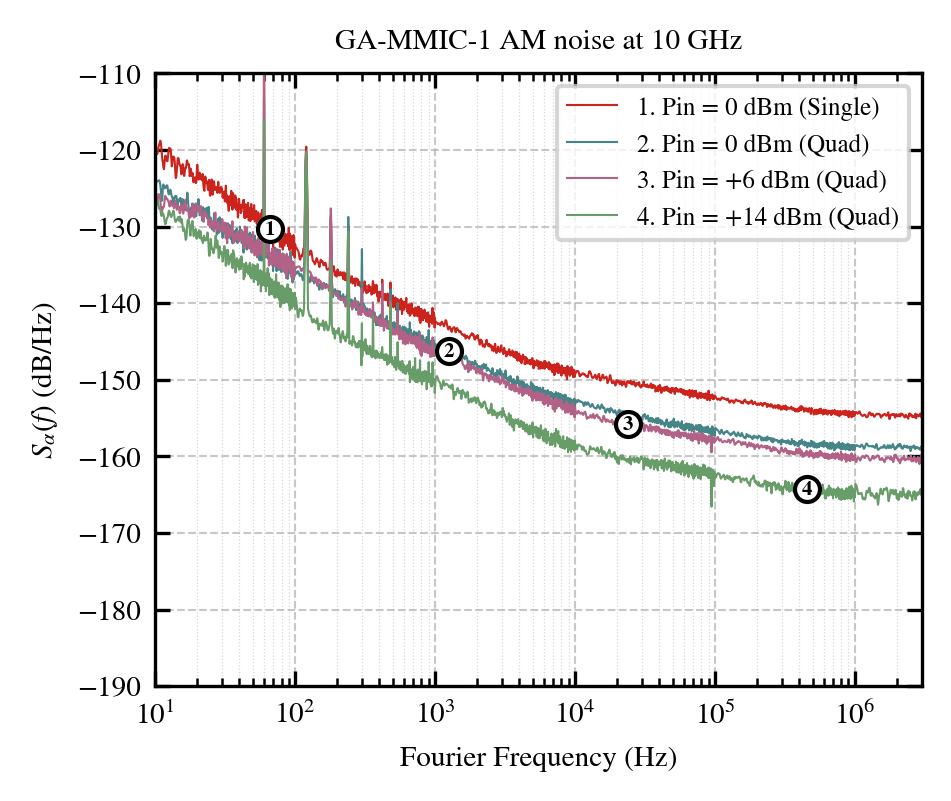}
  \caption{Residual AM noise of \ampC\ in single and quad
    configurations.}
  \label{fig:AM_ampC}
\end{figure}

Figures~\ref{fig:PM_ampC} and~\ref{fig:AM_ampC} show the PM and AM noise of \ampC\ in single and quad configurations. At 1~kHz offset, its flicker noise level is higher than that of \ampA\ by approximately 7~dB. The quad configuration provides the expected flicker improvement. The AM noise of \ampC\ exhibits the AM$>$PM behavior in both the flicker and thermal regions. 

\subsection{Effect of Voltage Regulator (\ampD, Single Amplifier)}
\label{sec:regulator}

\begin{figure}[htbp]
  \centering
  \includegraphics[width=\columnwidth]{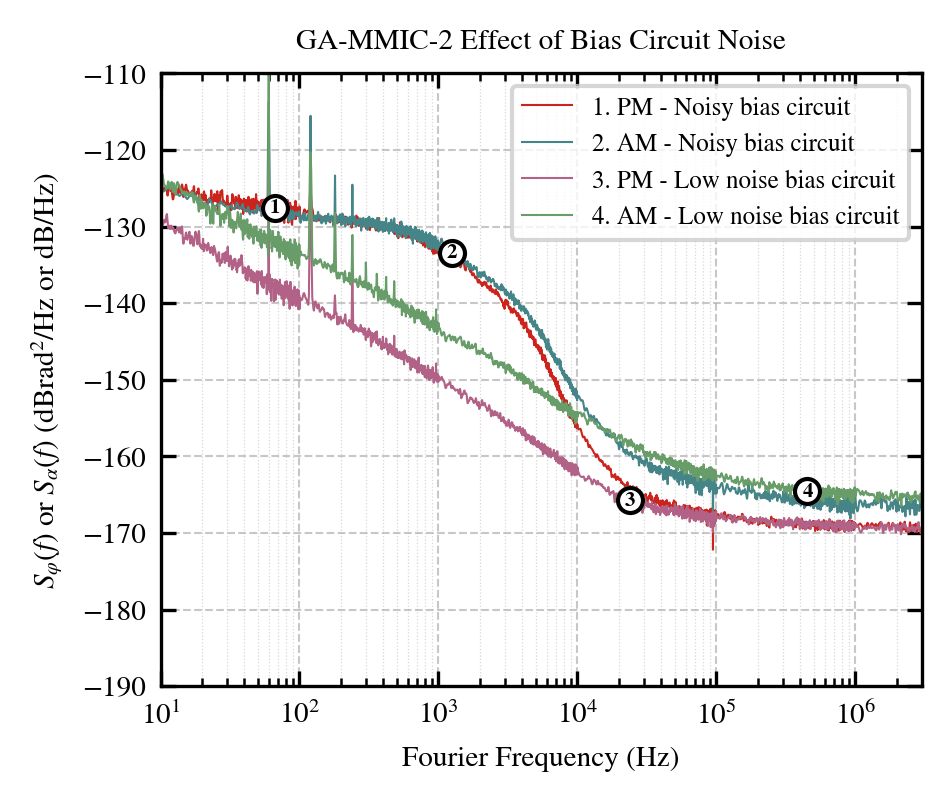}
  \caption{Effect of voltage regulator on residual AM and PM noise of the
    \ampD\ (single configuration) at 10~GHz. Commercial noisy bias module 
    vs.\ NIST custom low-noise bias supply.}
  \label{fig:Regulator_effect}
\end{figure}

Figure~\ref{fig:Regulator_effect} shows perhaps the most practically actionable result in the study, illustrating the residual PM and AM noise of \ampD\ under two bias supply conditions. The baseline measurement uses a commercially packaged \ampD\ amplifier with the supplied commercial bias module, which provides the drain voltage ($V_{DD}$) and gate voltage ($V_{GG}$) through low-dropout (LDO) regulators. The improved measurement replaces this module with a NIST custom-designed low-noise supply built around ultra-low-noise LDO regulators. The dominant noise contribution of the commercial bias module is its $V_{GG}$ regulator, whose voltage noise density at 1~kHz is $\approx 1700$ times higher than that of the corresponding regulator in the custom supply. This excessive bias noise degrades AM and PM by 10~dB and 17~dB, respectively, relative to nominal residual noise levels, and the degradation profile mirrors the noise spectrum of the commercial $V_{GG}$ regulator. Because the power supply-induced AM and PM noise levels are identical, the respective supply pushing conversion coefficients are likely equal, emphasizing the critical role of low-noise amplifier biasing.

\subsection{Remaining Devices}
\label{sec:appendix_devices}

\begin{figure}[htbp]
  \centering
  \includegraphics[width=\columnwidth]{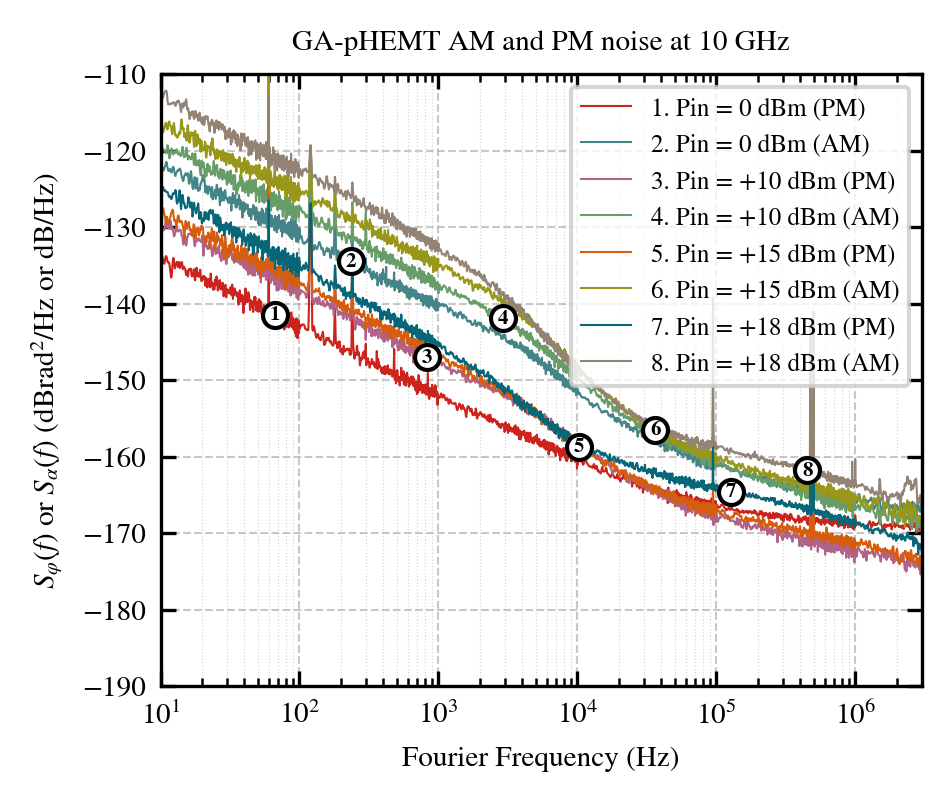}
  \caption{Residual AM and PM noise of \ampJ\ (GaAs pHEMT) in
    single configuration at various input power levels. The band-limited excess $1/f$ noise under 3~kHz offset may be a contribution of the manufacturer supplied bias circuit.}
  \label{fig:ampJ}
\end{figure}

\begin{figure}[htbp]
  \centering
  \includegraphics[width=\columnwidth]{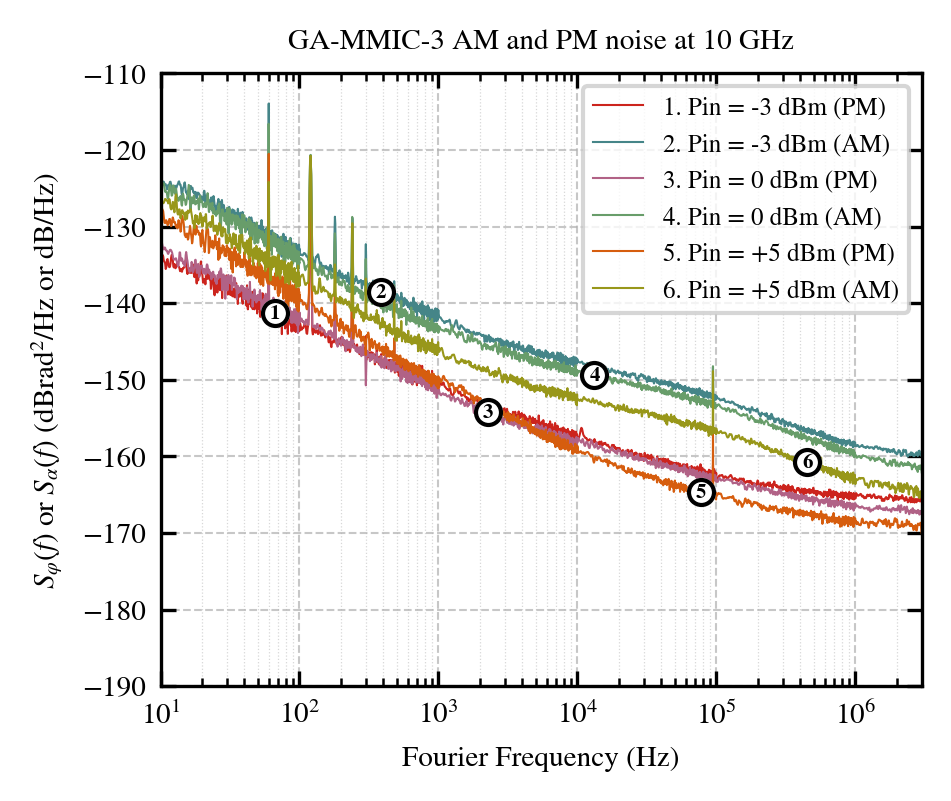}
  \caption{Residual AM and PM noise of \ampE\ in single
    configuration. An increase in input power results in a degradation of the residual flicker PM, whereas the AM performance exhibits a corresponding improvement.} 
  \label{fig:ampE}
\end{figure}

\begin{figure}[htbp]
  \centering
  \includegraphics[width=\columnwidth]{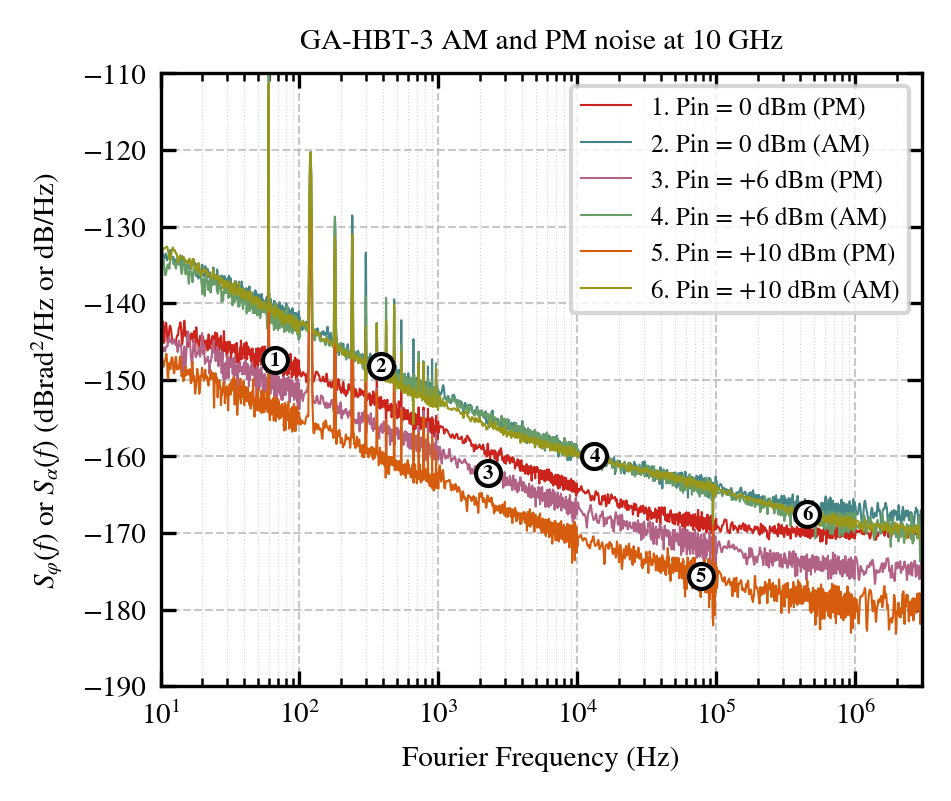}
  \caption{Residual AM and PM noise of \ampH\ in single
    configuration. At an input power of +10~dBm, AM noise is significantly higher than PM noise, exceeding it by more than 10~dB.}
  \label{fig:ampH}
\end{figure}

\begin{figure}[htbp]
  \centering
  \includegraphics[width=\columnwidth]{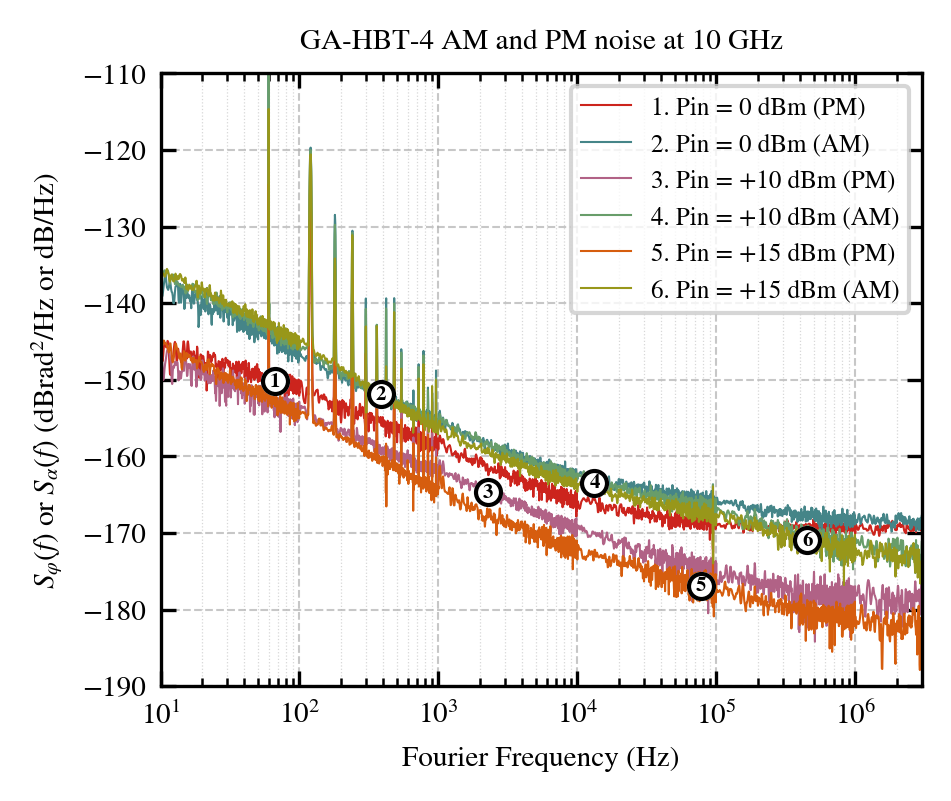}
  \caption{Residual AM and PM noise of \ampI\ in single
    configuration. The PM noise characteristics align closely with \ampH; however, \ampI\ shows a marginal improvement in AM noise, measuring a few decibels lower across the offset range.}
  \label{fig:ampI}
\end{figure}

\begin{figure}[htbp]
  \centering
  \includegraphics[width=\columnwidth]{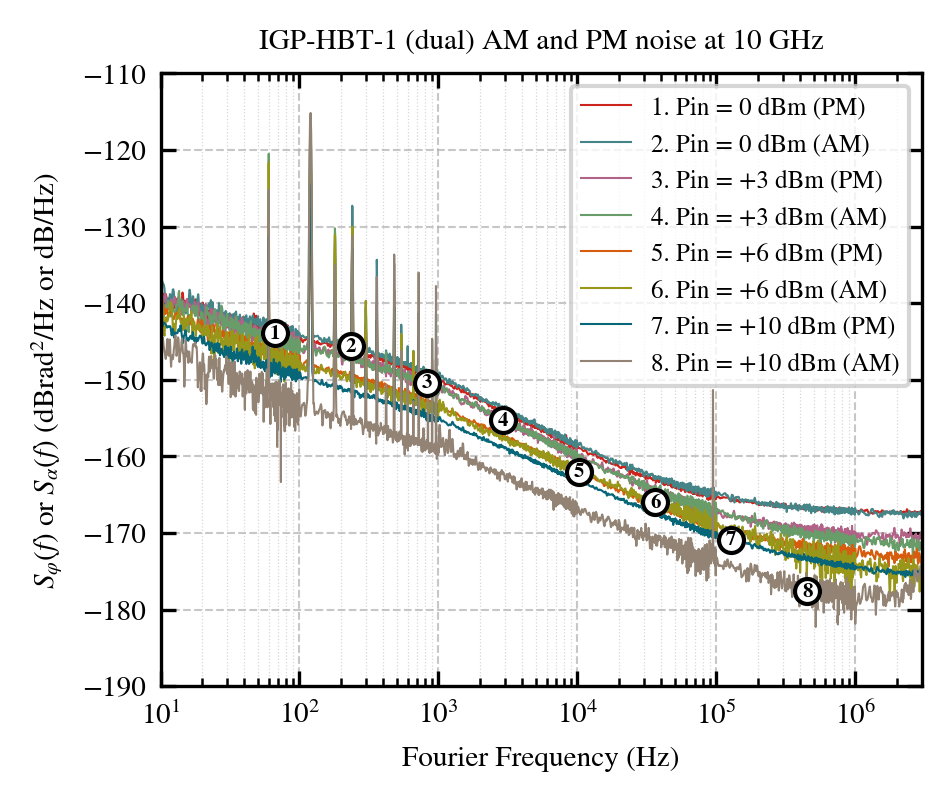}
  \caption{Residual AM and PM noise of \ampF\ in dual configuration. For this amplifier, the AM and PM noise are similar for any given input power level less than +10~dBm. AM noise measurement at +10~dBm is biased by a comparable reference oscillator AM noise.}
  \label{fig:ampF}
\end{figure}

\begin{figure}[htbp]
  \centering
  \includegraphics[width=\columnwidth]{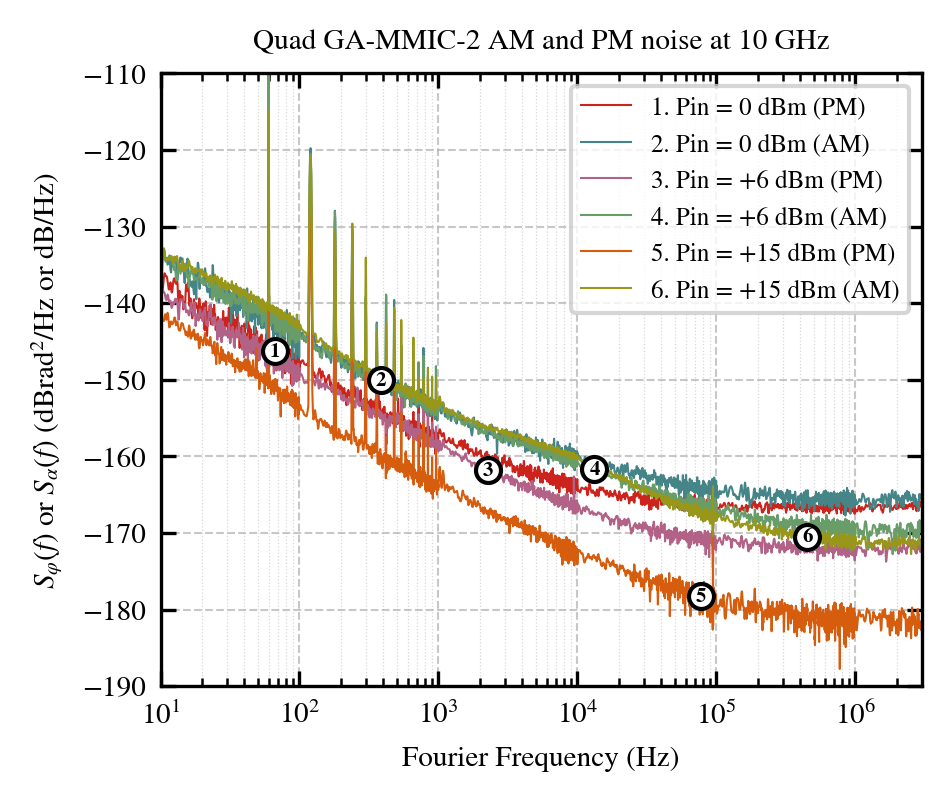}
  \caption{Residual AM and PM noise of \ampD\ in quad configuration. For an input power level of +15~dBm, the AM and PM noise performance is comparable to that of quad \ampA.}
  \label{fig:ampD_quad}
\end{figure}

\begin{figure}[htbp]
  \centering
  \includegraphics[width=\columnwidth]{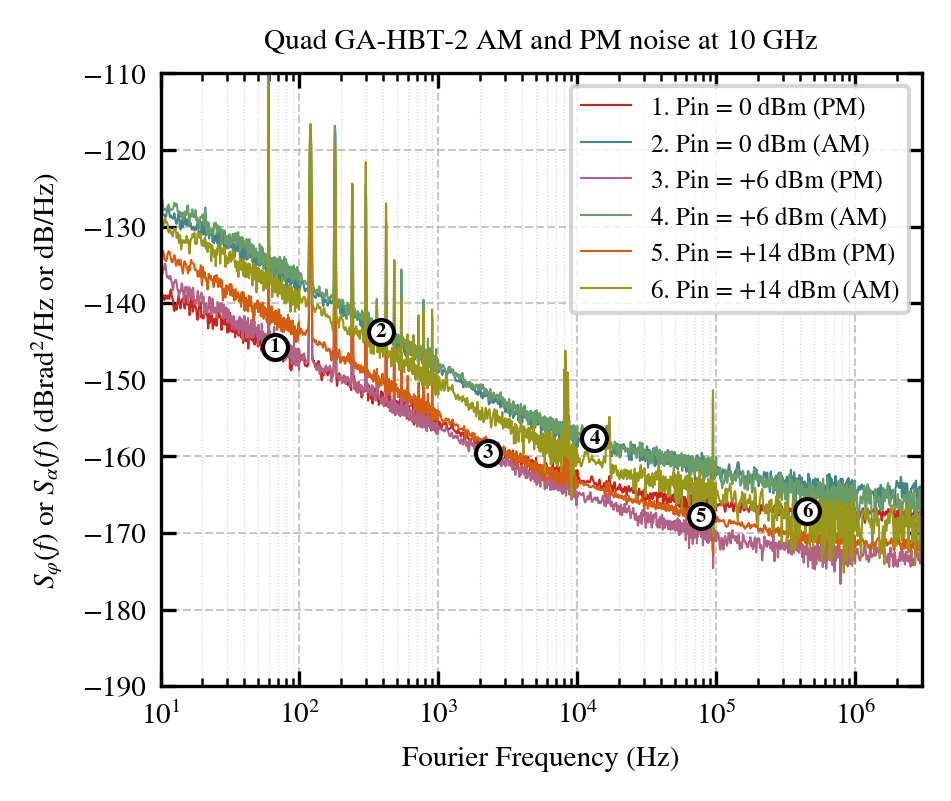}
  \caption{Residual AM and PM noise of \ampB\ in quad
    configuration. The noise is approximately 5~dB higher than that of the quad \ampA.}
  \label{fig:ampB_quad}
\end{figure}

\begin{figure}[htbp]
  \centering
  \includegraphics[width=\columnwidth]{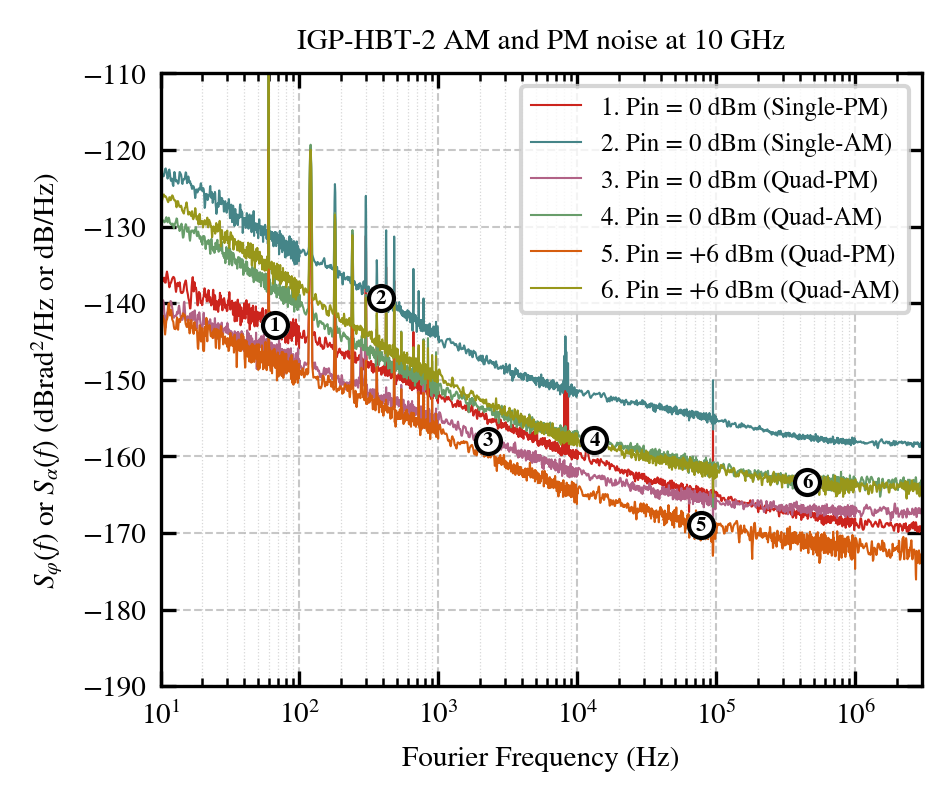}
  \caption{Residual AM and PM noise of \ampG\ in single and quad
    configurations.}
  \label{fig:ampG}
\end{figure}

\begin{figure}[htbp]
  \centering
  \includegraphics[width=\columnwidth]{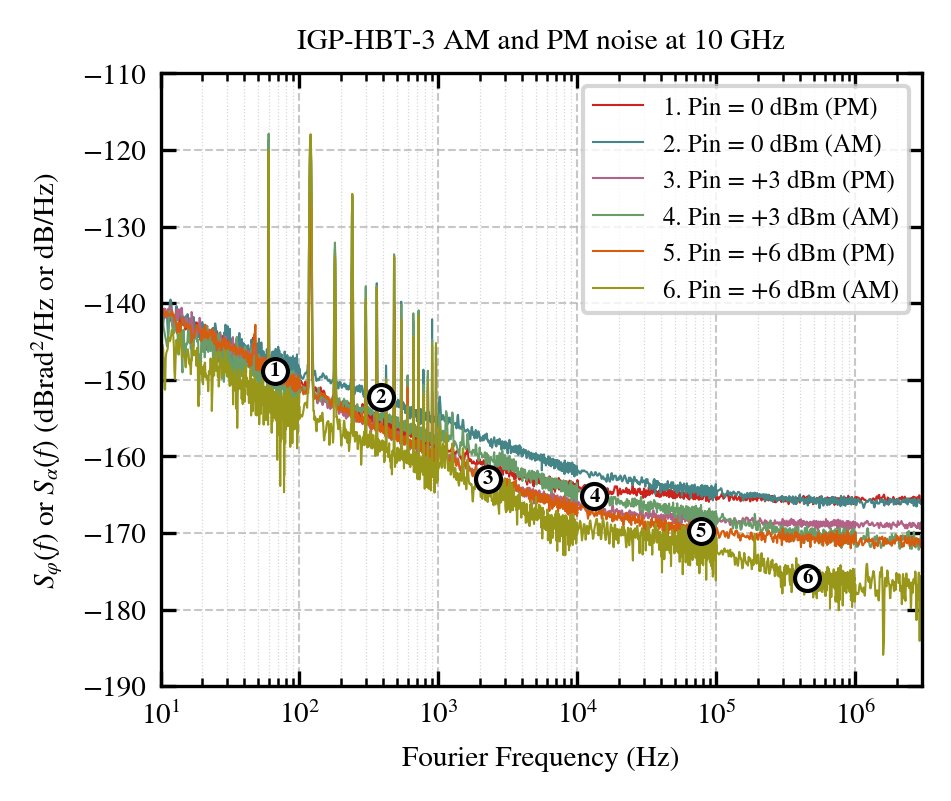}
  \caption{Residual AM and PM noise of \ampL\ in array
    configurations. In linear operation the white AM and PM noise both match the thermal limit $N_F k_B T / P_\mathrm{in}$; in deep compression the AM noise falls below this value. AM noise measurement at +6~dBm is biased by a comparable reference oscillator AM noise.}
  \label{fig:ampL}
\end{figure}

Figures~\ref{fig:ampJ}--\ref{fig:ampL} show the residual AM and PM noise for the remaining devices tested in this study. Key observations follow.

\textbf{\ampJ:} This pHEMT amplifier is included as a process outlier. As a GaAs pHEMT distributed amplifier with P1dB~$= +28$~dBm, it operates at 0~dBm approximately 15~dB below its compression point, the largest headroom margin of any device in this study. Notably, for a pHEMT, the PM noise is still comparable to that of the mid-power HBT devices. For this device AM and PM noise degrades as it enters compression (Fig.~\ref{fig:ampJ}). Also, excess $1/f$ noise near the carrier may be attributable to the commercial bias circuitry.

\textbf{\ampE:} Consistent with the other devices of the same product family (\ampC, \ampD), \ampE\ shows AM noise above PM noise in the flicker as well as thermal regions (Fig.~\ref{fig:ampE}). Its gain of $\approx 18$~dB is the highest of the three packaged MMICs tested. While entering saturation leads to an improvement in AM noise, it causes a degradation in PM noise. Within the flicker region, the AM and PM noise characteristics of a single \ampE\ are comparable to those of \ampC\ and \ampD\ (Figs. \ref{fig:Multi_Amp_PM} and \ref{fig:Multi_Amp_AM}).

\textbf{\ampH\ and \ampI:} These obsolete hermetic modules were characterized in single configuration (Figs.~\ref{fig:ampH} and \ref{fig:ampI}). \ampI, for instance, exhibits a phase noise of less than $-163$~dBrad$^2$/Hz at 1~kHz offset, measured at $P_\mathrm{sat}$ at 10~GHz (Fig. \ref{fig:ampI}). For these amplifiers, PM improves with compression while AM noise stays relatively constant.  Their noise is comparable to that of the \ampA\ quad amplifier.

\textbf{\ampF\ (Dual) array:} The \ampF\ unit examined in this study is configured as a dual-amplifier array, utilizing a shared voltage regulator to bias both amplifiers in the array. In its linear operating region, this array demonstrates similar AM and PM noise levels, with both improving as input power increases. However, as the device enters deeper compression with higher input power, the AM noise performance begins to surpass PM noise, eventually approaching the reference oscillator's AM noise (Fig.~\ref{fig:ampF}).

\textbf{\ampD\ (Quad):} \ampD\ in quad configuration is presented in Fig.~\ref{fig:ampD_quad}. It can produce almost 0.5~W of output power and its PM noise improves significantly in compression approaching $S_\varphi(100~\text{kHz})=-180~\text{dBrad}^2/\text{Hz}$. The flicker AM noise is independent of input drive level or compression.

\textbf{\ampB\ (Quad):} Figure~\ref{fig:ampB_quad} shows the AM and PM noise of \ampB\ in quad configuration. \ampB\ is designed for the 6~GHz to 12~GHz band specifically, in contrast to the broadband (2~GHz to 30~GHz) \ampA. Despite its narrower operating range, its quad noise is approximately 5~dB higher than that of the quad \ampA\ at a 1~kHz offset. 

The expected 6 dB flicker improvement for AM and PM noise is not fully met by the quad configuration unless the amplifiers are operating in saturation. However, saturation leads to degraded flicker PM but improved flicker AM noise (by a few decibels).

\textbf{\ampG\ (Single and Quad):} This InGaP HBT distributed amplifier in single and quad configuration is presented in Fig.~\ref{fig:ampG}. The \ampG\ quad configuration achieves $\approx 6$~dB improvement in AM noise over the single. Between 100~Hz and 1~kHz offset frequencies, this difference is approximately 7~dB, exceeding the ideal 6 dB improvement due to unit-to-unit performance variation. The close-to-the-carrier noise is comparable to the \ampC\ quad at comparable operating points. At a power level of 0~dBm, the PM thermal noise of the quad configuration is higher than that of the single amplifier, which suggests excess power splitter insertion loss.

\textbf{\ampL\ (Octa) array:} This discontinued amplifier consists of an array of eight parallel, dual-cascaded InGaP HBT amplifiers. It draws 27~W of DC power. The corresponding PM and AM noise characteristics are shown in Fig. \ref{fig:ampL}. Notably, its flicker PM noise is comparable to \ampI\ and the quad \ampA, while its AM noise aligns closely with that of the dual \ampF. This similarity in AM noise suggests that a higher-order array of \ampF\ units could potentially achieve the AM noise level of \ampL.

\section{Conclusion}
\label{sec:conclusion}

This study has characterized the residual AM and PM noise of twelve X-band, mostly HBT, amplifier designs at 10~GHz, identified by technology and package class, in single, dual, quad, and octa power-summed configurations. The principal findings are:

\begin{enumerate}
  \item In a quad power-summed configuration, the discrete-die HBT amplifier \ampA\ reaches below $-160$~dBrad$^2$/Hz residual PM noise and approximately $-155$~dB/Hz AM noise at 1~kHz offset, independent of input power, meeting the requirements of the target frequency synthesis application. The discontinued \ampI, \ampH, and \ampL\ modules measured residual PM noise between $-156$~dBrad$^2$/Hz and $-159$~dBrad$^2$/Hz at 1~kHz offset (Table~\ref{tab:dut}).
 
  \item \ampL, \ampI, \ampH, and \ampF\ measured AM noise between $-149$~dB/Hz and $-155$~dB/Hz at 1~kHz offset (Table~\ref{tab:dut}).
  Several devices show AM noise 5~dB to 10~dB above their own PM noise. This runs counter to the common system-budget assumption that AM and PM noise are equal in the thermal region, and it matters because AM noise enters the phase noise budget through AM-to-PM conversion at downstream nonlinear stages. However, under saturated conditions, \ampL\ and \ampF\ exhibit a reversal with AM noise falling below their PM noise.
   
  \item The $10\log_{10}(n)$ array improvement is approached but not fully reached: against the ideal 6~dB for a quad configuration. We typically observed 3~dB to 5~dB, limited primarily by excess splitter insertion loss, with unit-to-unit performance variation and impedance match also contributing. Typically, the performance of a power summed amplifier improves upon the noise of the individual amplifiers in the flicker region. For amplifiers with excess AM noise, the improvement extends into the white noise region as well because this excess AM noise is not additive in nature. 

  \item Both AM and PM noise depend on input drive level, but the trends are device specific (Figs.~\ref{fig:ampJ},~\ref{fig:ampH} and~\ref{fig:ampF}). In the \ampJ\ both components increase with input power; \ampH\ shows improving PM noise while its AM noise remains constant; and the dual \ampF\ improves in both AM and PM as input power increases.

  \item The voltage regulator is a first-order noise parameter, not a secondary effect. Replacing the supplied commercial bias module with a NIST low-noise supply (Sec.~\ref{sec:regulator}) reduces both the PM and AM noise of \ampD\ by more than 10~dB at 1~kHz offset. The dominant contribution originates from the $V_{GG}$ rail, where the supplied regulator's noise density exceeds that of the low-noise replacement by three orders of magnitude. This is critical because $V_{GG}$ couples directly into the bias current. For any application requiring noise below $-150$~dBrad$^2$/Hz at sub-kilohertz offsets, the choice of voltage regulator is as consequential as the choice of amplifier, and the lowest-noise regulator should be allocated to the $V_{GG}$ rail first.

  \item No commercial manufacturer currently specifies residual AM noise on a standard datasheet. This study reports measured X-band HBT amplifier AM and PM noise organized by device technology and topology, and is intended to serve as a reference for noise budgeting in precision frequency synthesis, radar, and timing applications.
 
\end{enumerate}

\section*{Acknowledgments}

The authors thank Benjamin Pera, Franklyn Quinlan, and Lindsay Sonderhouse for their careful reading and helpful feedback on this manuscript. This work is a contribution of the National Institute of Standards and Technology, an agency of the U.S.\ Government, and is not subject to U.S.\ copyright.

\section*{Disclaimer}

In accordance with NIST policy, the commercial devices characterized in this work are identified only by semiconductor technology and package class. No manufacturer or part number is given, the results are not a ranking, and no endorsement of any commercial product by NIST is implied.

\appendix[Analytical Model of Residual AM and PM Noise Measurement]
\label{app:residual_noise}
\setcounter{equation}{0}
\renewcommand{\theequation}{A\arabic{equation}}

This appendix demonstrates the AM$\leftrightarrow$PM leakage paths in
mixer-based residual noise measurements by use of a minimal
arctangent model of mixer port saturation, chosen over more
complete models \cite{cann_nonlinearity_1980,maas_nonlinear_2003}
for mathematical simplicity. The phase-to-voltage and amplitude
conversion factors are treated as calibration constants, and the
results are expressed in terms of the port overdrive ratios and
the phase error $\delta\theta$. A memoryless odd nonlinearity
transmits carrier phase exactly and therefore contains no
intrinsic AM-to-PM conversion. Physical mixers possess additional mechanisms, notably envelope-dependent junction capacitance and drive-dependent self-bias that convert AM to PM even at exact quadrature \cite{maas_nonlinear_2003}. For simplicity, this analysis omits these additional non-linear effects. Consequently, the leakage terms derived here represent lower bounds only due to the quadrature phase error.

The reference oscillator output voltage after the power splitter
(Fig.~\ref{fig:measurement_setup}) is modeled as a carrier with
small fractional amplitude fluctuations, $\alpha_s$, and phase
fluctuations, $\varphi_s$,
\begin{equation}
  v(t) = A\bigl[1 + \alpha_s(t)\bigr]
         \cos\bigl[2\pi f_0 t + \varphi_s(t)\bigr],
  \label{eq:source}
\end{equation}
with $|\alpha_s| \ll 1$ and $|\varphi_s| \ll 1$\,rad. Here, $A$ is the nominal peak amplitude and $f_0$ is the nominal frequency of the oscillator. After passing through the DUT, the signal acquires the additional residual fluctuations $\alpha_d(t)$ and $\varphi_d(t)$,
\begin{equation}
  v_1(t) = A_1\bigl[1 + \alpha_s(t) + \alpha_d(t)\bigr]
            \cos\bigl[2\pi f_0 t + \varphi_s(t)
            + \varphi_d(t)\bigr].
  \label{eq:dut_arm}
\end{equation}
The reference arm delivers a copy of the source with a phase
offset $\theta$ set by the quadrature phase shifter,
\begin{equation}
  v_2(t) = A_2\bigl[1 + \alpha_s(t)\bigr]
            \cos\bigl[2\pi f_0 t + \varphi_s(t) + \theta\bigr].
  \label{eq:ref_arm}
\end{equation}
Both arms carry the identical source fluctuations $\alpha_s(t)$
and $\varphi_s(t)$; we assume a delay-matched system throughout.

The saturating voltage response of a mixer port to an input
voltage $x$ is modeled as
\begin{equation}
  F(x) = \frac{2V_f}{\pi}
         \arctan\!\left(\frac{\pi x}{2V_f}\right),
  \label{eq:sat_model}
\end{equation}
normalized to unit small-signal slope and saturating
asymptotically to the hard limits $\pm V_f$, where $V_f$ is the saturation voltage of the port. Each port is
characterized by its overdrive ratio,
\begin{equation}
  \xi = \frac{A}{V_f},
  \qquad
  \xi_\text{RF} = \frac{A_1}{V_{f,\text{RF}}},
  \qquad
  \xi_\text{LO} = \frac{A_2}{V_{f,\text{LO}}},
  \label{eq:xi_def}
\end{equation}
with $\xi \ll 1$ corresponding to linear operation and
$\xi \gg 1$ to hard limiting. Because $F$ is odd and memoryless, the carrier zero crossings, and hence the instantaneous phase,
are transmitted exactly at any drive level, whereas the envelope is not. The detector response is dependent on the fundamental Fourier amplitude $a_1(\tilde{A})$ of the port output, evaluated at the quasi-static envelope $\tilde{A} = A[1+\alpha]$. Expanding
$a_1$ to first order in $\alpha$ identifies the fractional AM
transmission coefficient, $\eta$,
\begin{equation}
  a_1\bigl(A[1+\alpha]\bigr)
  \approx a_1(A)
  \left[1 +
  \underbrace{\frac{A\,a_1'(A)}
                   {a_1(A)}}_{\textstyle \eta}
  \,\alpha\right].
  \label{eq:eta_def}
\end{equation}
Evaluating $a_1$ for the model \eqref{eq:sat_model} gives the
closed form
\begin{equation}
  \eta(\xi)
  = \frac{1}{\sqrt{1+(\pi\xi/2)^2}},
  \label{eq:eta_closed}
\end{equation}
which approaches unity in linear operation, falls as $2/(\pi\xi)$
in hard limiting, and vanishes only for an ideal limiter.
Saturation therefore compresses, but never eliminates, the AM
transmitted to the detector.

The phase detector multiplies its two port signals and after low-pass filtering the product retains the difference-frequency term.
With the transmitted port envelopes written as $m_1(t)$ and
$m_2(t)$, the baseband output is
\begin{equation}
  v_\text{out}(t) = k\,
  m_1(t)\,m_2(t)\,
  \cos\bigl[\varphi_d(t) - \theta\bigr],
  \label{eq:detector}
\end{equation}
where $k$ is the operating point dependent conversion constant obtained by calibration
($k_\varphi$ in V/rad for the PM configuration, $k_\alpha$ in
volts for the AM configuration), $m_1 = 1 + \eta(\xi_\text{RF})(\alpha_s+\alpha_d)$ and
$m_2 = 1 + \eta(\xi_\text{LO})\alpha_s$. The
source PM noise $\varphi_s$ enters both ports identically and
cancels exactly in \eqref{eq:detector}, which is the fundamental
advantage of the homodyne measurement topology. In practice the
phase setting contains a small phase error,
$\theta = \theta_0 + \delta\theta$ with
$|\delta\theta| \ll 1$\,rad.

In the PM configuration both ports are saturated, so
setting $\theta_0 = \pi/2$ in \eqref{eq:detector} and expanding to first
order gives
\begin{equation}
  \begin{split}
  v_\text{out}^\text{PM}(t) = k_\varphi \bigl[
  \varphi_d(t)
  &- \eta(\xi_\text{RF})\,\alpha_d(t)\,\delta\theta\\
  &- \bigl(\eta(\xi_\text{RF})
     + \eta(\xi_\text{LO})\bigr)
     \alpha_s(t)\,\delta\theta \bigr].
  \end{split}
  \label{eq:pm_crosstalk}
\end{equation}
The DUT AM noise leaks only through the RF port, as indicated by
the second term in \eqref{eq:pm_crosstalk}. The source AM noise
leaks through both ports, and because both contributions are
driven by the same fluctuation they add coherently in amplitude.
For our simplified saturation model,  both incomplete saturation ($\eta \neq 0$) and a phase error ($\delta\theta \neq 0$) are required simultaneously for AM noise to contaminate the PM channel at first order.

In the AM configuration the LO port is saturated and the RF port
is linear ($\eta \to 1$), so $m_1 = 1 + \alpha_s + \alpha_d$.
Setting $\theta_0 = 0$ and expanding similarly gives
\begin{equation}
   \begin{split}
  v_\text{out}^\text{AM}(t) = k_\alpha \bigl[
  1
  + \bigl(1+\eta(\xi_\text{LO})\bigr)\alpha_s(t)\\
  + \alpha_d(t)
  + \varphi_d(t)\,\delta\theta \bigr].
   \end{split}
  \label{eq:am_crosstalk}
\end{equation}
The PM-to-AM leakage term $\varphi_d\,\delta\theta$ carries no
$\eta$ factor, the in-phase operating point sits at the maximum
of the detector characteristic rather than at a zero crossing.
Therefore, a small phase error introduces PM-to-AM leakage
regardless of the saturation state of the ports. The source AM
noise enters with coefficient $1+\eta(\xi_\text{LO})$, as
indicated by the second term in \eqref{eq:am_crosstalk}. Unlike PM measurements 
there is no differential cancellation mechanism for source AM noise.

Collecting the first-order contributions, and for simplicity assuming all noise terms are uncorrelated, the measured
one-sided double sideband (DSB) noise PSDs are 
\begin{equation}
    \begin{split}
      S_\varphi^\text{meas}(f) =
        S_{\varphi_d}(f)
        &+ \delta\theta^2\,
          \eta^2(\xi_\text{RF})
          \,S_{\alpha_d}(f)\\
        &+ \delta\theta^2
          \bigl[\eta(\xi_\text{RF})
          + \eta(\xi_\text{LO})\bigr]^2
          \,S_{\alpha_s}(f),
    \end{split}
    \label{eq:pm_total}
\end{equation}
\begin{equation}
    \begin{split}
      S_\alpha^\text{meas}(f) = S_{\alpha_d}(f)
        &+ \bigl(1 + \eta(\xi_\text{LO})\bigr)^2
          S_{\alpha_s}(f)\\
        &+ \delta\theta^2\,S_{\varphi_d}(f),
    \end{split}
    \label{eq:am_total}
\end{equation}
with $\eta(\xi)$ from \eqref{eq:eta_closed} and the port
overdrive ratios from \eqref{eq:xi_def}. Even in the
hard-limiting limit $\eta(\xi_\text{LO}) \to 0$, the source AM
term in \eqref{eq:am_total} retains unit coefficient, so valid
residual AM characterization requires
\begin{equation}
  S_{\alpha_s}(f) \ll S_{\alpha_d}(f)
  \label{eq:am_requirement}
\end{equation}
across the offset range of interest, a constraint with no
analogue in the PM measurement.

 Equations \eqref{eq:pm_total} and \eqref{eq:am_total} constitute the principal results of this appendix, leading to four primary conclusions. First, the source PM noise is absent from both expressions at all saturation
levels.
Second, AM noise contaminates the
PM channel when both $\eta(\xi) \neq 0$ and
$\delta\theta \neq 0$ simultaneously. Third, PM noise contaminates the
AM channel whenever $\delta\theta \neq 0$, regardless
of saturation depth. Finally, even under ideal conditions, the AM noise measurement retains the source
term $S_{\alpha_s}$ through the linear RF port. This highlights a fundamental asymmetry between the two modes: source PM cancels differentially, whereas source AM does not. Consequently, a valid residual AM
noise measurement using Fig.~\ref{fig:measurement_setup} strictly requires
$S_{\alpha_s} \ll S_{\alpha_d}$.
Alternative measurement schemes do exist that utilize differential architectures to effectively suppress reference-oscillator AM noise~\cite{rubiola_measurement_2005,pomponio_flicker_2026}.


\end{document}